\documentclass[a4paper,12pt]{article}
\usepackage[flushmargin]{footmisc}
\usepackage{amsmath}
\usepackage{amsfonts}
\usepackage[latin1]{inputenc}
\usepackage{rotating,graphics,eurosym}
\usepackage{longtable,lscape}
\usepackage[english]{babel}
\usepackage{pgf,pgfarrows,pgfnodes,pgfautomata,pgfheaps,pgfshade}
\usepackage{float}
\usepackage[comma]{natbib}
\usepackage{setspace}
\usepackage{multirow}
\usepackage[dvips]{epsfig}
\usepackage[dvips]{graphicx}
\usepackage{amssymb}
\usepackage{threeparttable}
\usepackage{units}
\usepackage{caption}
\usepackage{tabularx}
\usepackage{subfigure}
\DeclareGraphicsExtensions{.eps .jpg} 

 \paperheight29,7cm \paperwidth21cm

\begin{document}

\pagenumbering{arabic} \setlength{\unitlength}{1cm}
\pagestyle{plain}  \normalsize
\title{Identifying causal channels of policy reforms with multiple treatments and different types of selection\footnote{This study is part of the project ``Regional Allocation Intensities, Effectiveness and Reform Effects of Training Vouchers in Active Labor Market Policies'', IAB project 1155. This is a joint project of the Institute for Employment Research (IAB) and the University of Freiburg. We gratefully acknowledge financial and material support from the IAB. The paper was presented at ESPE in Aarhus, CAFE Workshop in B$\o$rkop, SOLE in Washington, EALE in Ljubljana, Joint Research Centre of the European Commission, Centre for European Economic Research, and the University of Bern. We thank participants for helpful comments, in particular Hugo Bodory, Bernd Fitzenberger, Hans Fricke, Michael Lechner, Michael Knaus, Thomas Kruppe, Marie Paul, and Gesine Stephan. We are particularly grateful for detailed comments and remarks from Conny Wunsch. Furthermore, we thank two anonymous referees. The usual disclaimer applies. Correspondence: annabelle.doerr@berkeley.edu, anthony.strittmatter@unisg.ch}}

\author{Annabelle Doerr \vspace{-0.3cm}\\ 
\small{UC Berkeley} \vspace{-0.3cm}\\ 
\small{University of Basel}\\ 
\and Anthony Strittmatter \vspace{-0.3cm} \\
\small{University of St. Gallen}}

\date{\today}
\maketitle
\vspace{-0.6cm}
\begin{abstract}\small
We study the identification of channels of policy reforms with multiple treatments and different types of selection for each treatment. We disentangle reform effects into policy effects, selection effects, and time effects under the assumption of conditional independence, common trends, and an additional exclusion restriction on the non-treated. Furthermore, we show the identification of direct- and indirect policy effects after imposing additional sequential conditional independence assumptions on mediating variables. We illustrate the approach using the German reform of the allocation system of vocational training for unemployed persons. The reform changed the allocation of training from a mandatory system to a voluntary voucher system. Simultaneously, the selection criteria for participants changed, and the reform altered the composition of course types. We consider the course composition as a mediator of the policy reform. We show that the empirical evidence from previous studies reverses when considering the course composition. This has important implications for policy conclusions. \\
\end{abstract}

\noindent \small
%Word count: 9018 + 5.5 pages Figures and Tables\\
JEL-Classification:  C21, J68, H43

\normalsize\maketitle
\doublespacing
\newpage

\section{Introduction}
%The gold standard of evidence-based policy making is the evaluation of policy instruments using randomised control trials prior to a large-scale implementation. In practice, randomized control trials are only rarely implemented because they are costly, time demanding, and may receive minimal social acceptance. In such cases, policy makers and researchers have to deal with the non-random nature of policy instruments. 

A popular approach to evaluate the effectiveness of policy reforms in quasi experimental settings is the Difference-in-Differences (DiD) method. The baseline version of DiD requires to observe one treatment and one control group. Both groups are untreated before the policy reform. After the reform the treatment group receives the treatment while the control group remains untreated. The effectiveness of the reform can be estimated by comparing the differences in outcomes of both groups before and after the reform implementation. This comparison will lead to unbiased estimates under the common trend assumption, i.e., when the outcomes of both groups would have developed parallel to each other in the absence of the policy reform.

Often the evaluation of policy instruments does not work that simply. For this reason, the literature proposes several extensions of the baseline DiD method. Some studies impose conditional independence assumptions to account for selection into treatment based on observable characteristics \citep[e.g.,][]{aba05,heck97,lech10}.  Other studies consider multiple treatments \citep[e.g.,][]{fri17}. This capture situations in which the policy of interest is the reform of an existing policy instrument, for example an increase of treatment intensity, instead of the implementation of a new instrument. There are studies that combine both extensions \citep[e.g.,][]{fel14,hav11a,hav11b}. Some studies even consider multiple treatments and different types of selection for each treatment \citep[e.g.,][]{car05,ri13}. However, the results from these studies are not formally identified without the implementation of a structural model for the specific policy question. 

As the first contribution of our paper, we formally show how policy reforms can be non-parametrically decomposed into effects of changing the policy instrument and selection effects, as well as other time changing factors such as business cycle effects. We mainly rely on conditional independence and common trend assumptions. We highlight that an additional exclusion restriction on the untreated is sufficient to identify the policy and time effects which is not recognised by the previous literature. The imposed assumptions are necessary and sufficient to achieve additive separability, which is, for example, imposed by \cite{ri13}. 

Second, we focus on the direct and indirect channels through which changes in a policy instrument may unfold their effects. Relying on mediation analysis \citep[see, e.g.,][for a review about mediation analysis]{huber17}, we are the first who explore direct and indirect effects of a quasi-experimental policy reform. The existing approaches of the mediation literature investigate direct and indirect treatment instead of policy reform effects \citep[see, e.g.,][]{flo09,hub15,ima10,ima11,peter06,weele09}. Closely related is the study by \cite{hub18}, who use common trend assumptions to identify direct and indirect policy effects.\footnote{Relatedly, \cite{hub19} use a changes-in-changes framework to identify direct and indirect channels.} In contrast, we rely on common trend assumptions to identify the policy effects and impose additional sequential independence assumptions on the mediators to identify the direct and indirect effects of the policy reform on the outcome of interest. 

Third, we illustrate our approach using a large-scale reform of the allocation system of unemployed individuals to vocational training in Germany. The reform replaced the existing mandatory allocation system with a voucher allocation system. The \emph{voucher system} offers voluntary participation and participants have (some) influence on the course choice \citep[see detailed discussion in][]{doe16}. Under the \emph{mandatory system}, participation was compulsory and caseworkers in local employment agencies allocated participants to specific courses. Additionally, the reform changed the criteria for selecting unemployed persons into training programmes. Under the pre-reform system, caseworkers assigned training based on subjective criteria, whereas the new selection rule focuses on predicted future employment outcomes. Caseworkers were incentivised to select unemployed persons with an expected re-employment probability of at least 70\% within six months after the end of training. Accordingly, this reform offers a setting in which the overall reform effect is a composition of time effects, selection effects and the policy effects of interest - in this case the effects of changing the allocation of vocational training from a mandatory system to a voucher system.

Furthermore, this reform provides an illustrative example of a situation in which the policy change may result in direct and indirect effect on the outcome. First, voluntary participation might increase the motivation of participants compared to participants who are assigned in a mandatory system (direct effect). Second, unemployed persons and caseworkers might choose different types of courses (indirect effect). The existing literature on training allocation systems mainly focuses on the effects of different degrees of participants freedom of course choice \citep[see, e.g., the surveys][]{mc16, to16, st16}.\footnote{For example, \cite{pr11} provide experimental evidence for the relative effectiveness of different degrees of participants' influence on the course choice under voluntary participation in training. They find that increasing participants' course choices has no effects on his or her re-employment probability and negative effects on his or her earnings.} However, it is unexplored whether training is more effective under voluntary or mandatory participation, net of the course composition that might change under different allocation systems. Our study also contributes to close this research gap.

We build on the work of \cite{ri13}, who investigate the same reform. They disentangle the effects of the reform of the allocation systems from the changing selection criteria and find positive but mostly insignificant short-term effects of the voucher reform. We replicate their results using a larger data set and an efficient estimation method. \cite{ri13} observe 1,319 training participants after the reform and match control observation by single-nearest-neighbour matching. In contrast, we obtained administrative data from the Federal Employment Agency of Germany, which contain the population of vocational training participants during the years 2001-2004. Our evaluation sample consists of more than 26,000 training participants in each time period. We apply the doubly robust and locally efficient auxiliary-to-study tilting estimator proposed in \cite{gr16}.

Furthermore, our data allows us to consider long-term effects over a time period of more than seven years after programme entry \citep[in contrast to 1.5 years in][]{ri13}. Qualitatively we confirm the findings for the time period under consideration in \cite{ri13}. Our results suggest positive effects of the reform of the allocation system in the short-term. Moreover, we find that the reform of the allocation system reduces the re-employment probabilities between the first and second year after the start of training. After three years, the effects turn positive and remain on an approximately stable level until seven years after the training started. This suggests that it is crucial to consider long-term reform effects.

In contrast to \cite{ri13}, we consider the type and duration of training as mediators, i.e., intermediate outcomes on the causal path of the assignment system to the individual labour market outcomes. Our results show that the short-term positive effects of the reform are mainly driven by a different composition of training course types and durations after the reform. More individuals participate in shorter courses in the post-reform period which leads to an improvement of labour market outcomes in the short-term but not in the long-term. This is almost a mechanical effect, because participants in courses with short durations are distracted from intensive job search for a shorter time period. 

This is an example of Manski's (1997) mixing problem in programme evaluations. Treatment variation occurs because participant can self-select into different types and durations of training. This makes the evaluation of the treatment particularly complicated, because it is difficult to disentangle variations in the allocation system or treatment. \cite{man97} suggests an partial identification approach to address the mixing problem \citep[see also the discussion in][]{gun17}. We follow a different strategy and use a mediation analysis framework \cite[e.g.][]{ima11} to separate the effects of the voluntary allocation system from the variation in the types and durations of training. 

We are particularly interested in the effects of voluntary participation net of the effects from a changing course composition. We find negative employment effects during the first three years after programme entry. During the lock-in period the re-employment chances decrease by up to four percentage points. A possible explanation for this result is lower job search intensity under voluntary participation which may be explained by a higher motivation to attend and complete the courses. The effects tend to turn positive in the long-term. Possibly, unemployed individuals accumulate more human capital under the voluntary system than under the mandatory system, which pays off in the long-term. These results point out that causal channels largely affect the policy conclusions. From a policy maker perspective, voluntary participation should only be offered when the programmes' objective is a long-term investment in human capital. Mandatory assignment appears to be more successful in the short-term. Accordingly, this allocation system should be used when fast reintegration is the major programme goal. 

%Another interesting result of our study focuses on the selection criteria reform. We assess that the new selection criteria for programme participants are poorly designed. Under the new selection criteria, caseworkers have an incentive to allocate unemployed individuals with good labour market opportunities to programmes with shorter durations. This strategy helps caseworkers to conform with the 70\% rule but does not increase the efficiency of vocational training. 
%We could conclude that the reform had a positive impact on the effectiveness of training in the short-term that turns negative after one year and remain on a marginally positive level in the longer run. However, a carefully implemented analysis of indirect and direct effects of the reform reveals that the positive effects in the short-term are exclusively driven by a changing course composition. Net of the course composition, the reform from a mandatory to a voluntary allocation system only reveals small positive but close to zero effects in the long-run. 

The remainder of this paper is structured as follows. In the next section, we show the identification of the policy effect of a reform and its causal channels in a setting with multiple treatments and selection. We discuss the parameter of interest, identification, and estimation strategy. A detailed illustration of this approach using the example of the allocation reform of vocational training in Germany follows in Section \ref{sec3:training}. The final section concludes. Additional information is provided in Online Appendices \ref{app1}-\ref{app5}.

\section{Identification of reform effects and causal channels} \label{sec2:empirical}

\subsection{Parameters of interest} \label{sec21:parameter}
We define the parameter of interest within the potential outcome framework proposed by \cite{rub74}. We denote random variables by capital letters and realized values by small letters. Assume we have a random sample of individuals from a large population. For each individual in the sample, we observe the treatment state $D = d\in\{0,1\}$ which indicates whether the individual receives a treatment $D = 1$ or not $D = 0$. Furthermore, we assume that a reform of the policy instrument took place at some point in time. Let $T$ be an indicator for the time period that can take on the values $t\in\{0,1\}$ for the pre-reform or post-reform time period, respectively. Finally, we consider a policy system indicator $S = s\in\{b,a\}$ that is $b$ before the reform was implemented and $a$ afterwards.

We indicate the potential outcomes by $Y^d_t(s)$. They can be stratified into eight groups: $Y^1_0(b)$ and $Y^1_1(b)$ indicate the potential outcomes that would be observed if under pre-reform system treatment in the pre- or post-reform period, respectively. $Y^1_0(a)$ and $Y^1_1(a)$ are the potential outcomes under post-reform system treatment in the pre- or post-reform period. $Y^0_0(b)$ and $Y^0_1(b)$ are the potential outcomes under pre-reform system non-treatment before or after the reform. $Y^0_0(a)$ and $Y^0_1(a)$ are the analogous potential outcomes under post-reform system non-treatment in both time periods.

We only observe one potential outcome for each individual. We never observe  pre-reform system treatments after the reform took place ($Y^1_1(b)$, $Y^0_1(b)$). Similarly, we never observe the post-reform system treatments in the pre-reform period ($Y^1_0(a)$, $Y^0_0(a)$) because the post-reform policy system was implemented as part of the reform. The observed outcome equals

\begin{equation*}
Y = \sum\limits_{d\in\{0,1\}}\sum\limits_{t\in\{0,1\}}\sum\limits_{s\in\{b,a\}} G(d,t,s)Y_{t}^d(s),
\end{equation*}
where $G(d,t,s)$ is an indicator function with $G(d,t,s) = 1\{D=d,T=t,S=s\}$ for $d,t\in\{0,1\}$ and $s\in\{b,a\}$, which is the stable unit treatment value assumption (SUTVA) \citep[e.g.,][]{cox58}. 

In our application, $D$ specifies whether an unemployed individual participates in a vocational training programme and $S$ specifies the training allocation system before (mandatory system $m$) and after the reform (voucher system $v$). The outcome $Y$ measures different labour market outcomes. 

We are primary interest in the policy effect, i.e., the effect of the reform of the allocation into training from a mandatory to a voucher system. Policy effects are the expected difference of potential outcomes under the voucher and mandatory systems by holding treatment status and time period constant. In particular, we focus in our application on the policy effects under treatment in the post-reform period\footnote{Alternatively, the policy effect could be defined under non-treatment status or for the pre-reform period.}
\begin{align*}
 & \color{blue} \gamma^{p} = E[Y^1_1(v)-Y^1_1(m)|D=1,T=1].
\end{align*}

Consider the following thought experiment to clarify the interpretation of this policy effect: Compare the employment outcomes of training participants who receive a training voucher after the reform with the employment outcomes that they would obtain if they were mandatory assigned to a (potentially different) vocational training course after the reform. In the following, we show how to derive this parameter.

As a starting point, we consider the overall reform effect, i.e., the  comparison of the effectiveness of a policy instrument before and after the reform. In our application it is defined as the difference between the effectiveness of participating in training courses in the post- and pre-reform period,
\begin{align*}
 & \gamma^{ba} = E[Y^1_{1}(v) - Y^0_{1}(v)|D=1,T=1] - E[Y^1_{0}(m) - Y^0_{0}(m)|D=1,T=0]. 
\end{align*}

We show below how to decompose the overall effect into the selection effect, the time effect, and policy effect. It is often the case, that reforms of policy instruments also effect the selection of those who are treated with the policy instrument. In our example, part of the reform was the implementation of stricter selection criteria of participants. As a consequence, treated individuals before and after the reform may differ in their observed characteristics. The selection effect under the mandatory system in the pre-reform period can be formalised as
\begin{align*}
\gamma^{sel}= E[Y^1_{0}(m)-Y^0_{0}(m)|D=1,T=1]-E[Y^1_{0}(m)-Y^0_{0}(m)|D=1,T=0].
\end{align*}

The treated population may change before and after the reform, but the policy system and time effects are held constant. The following thought experiment may clarify the interpretation of the selection effect: Assign participants from the post-reform period to training in the pre-reform period. Then, compare them to actually observed participants in the pre-reform period.

Furthermore, the labour market outcomes of individuals could differ before and after the reform because of time effects even after controlling for treatment state and policy system. In our setting, it is likely that business cycle effects occur. In our application, we define the business cycle effects under the mandatory system for the treated population after the reform, which we formalise as 
\begin{align*}
	& \gamma^{bc1} = E[Y^{1}_1(m)-Y^{1}_0(m)|D=1,T=1] \mbox{ and}\\
		& \gamma^{bc0} = E[Y^{0}_1(m)-Y^{0}_0(m)|D=1,T=1].
\end{align*}
The parameters $\gamma^{bc1}$ defines business cycle effects under treatment in the mandatory system and  $\gamma^{bc0}$ defines business cycle effects under non-treatment in the mandatory system. In the following, we discuss the sufficient assumptions to identify the effects of interest. 

\subsection{Identification of reform effects}\label{sec22:ident}
The identification of the overall reform effect $\gamma^{ba}$ and selection effects $\gamma^{sel}$ from the joint distribution of random variables $(Y,G(d,t,s),X)$ can be achieved by controlling for a large set of $K$ confounding pre-treatment variables $X$ with support $\mathbb{X} \subseteq \mathbb{R}^K$ to account for the possibility of selection into treatment based on observed characteristics. 

\bigskip
\noindent \textbf{Assumption 1a} \emph{(Conditional Mean Independence)}\\
For all $d,d',t,t' \in \{0,1\}$, $s\in \{m,v\}$ and $x \in \mathbb{X}$,
\begin{align*}
	& E[Y^d_t(s)|D=d', T=t', X=x]  = E[Y^d_t(s)|D=d, T=t, X=x]
\end{align*}
and all necessary moments exist.
\bigskip 

This assumption implies that the expected potential outcomes are independent of the treatment $D$ and time period $T$ after controlling for the pre-treatment control variables $X$. All confounding variables, which jointly influence the expected potential outcomes and treatment status must be included in the vector $X$. Note that Assumption 1a also includes a time dimension, i.e., we assume that individuals being treated in $t=1$ would have the same expected potential outcomes as treated individuals in $t=0$ if they were treated under the pre-reform policy system before the reform (conditional on $X$). This assumptions holds if those treated before and after the reform do not differ systematically in unobserved characteristics that influence both the treatment probability and potential outcomes. 

\bigskip
\noindent \textbf{Assumption 2a} \emph{(Support)}. \\
\begin{equation*}
	0< Pr(G(d,t,s)=1|X=x)< 1 \ \  \forall \ \  d,t\in\{0,1\}
\end{equation*}
for the subpopulation with $G(d',t',s)=1 \ \  \forall \ \ d',t'\in\{0,1\}$.
\bigskip 

Assumption 2a requires overlap in the propensity score distributions of the different sub-populations, which can be tested in the data \citep[see the discussion in][]{lech13}. 

Under Assumptions 1a and 2a, for all $d,d',t,t' \in \{0,1\}$ and $s\in\{m,v\}$
\begin{equation} \label{proof1}
E[Y_{t}^d(s)|D=d', T=t'] = E \left[\frac{p_{d',t',s}(X)}{p_{d',t',s}\cdot p_{d,t,s}(X)}G(d,t,s)Y\right],
\end{equation}
is identified from observed data on the joint distribution of
$(Y,G(d,t,s),G(d',t',s),X)$, with $p_{k,l,s}(x)=Pr(G(k,l,s)=1|X=x)$ and $p_{k,l,s}=Pr(G(k,l,s)=1)$ for $k\in\{d,d'\}$ and $l \in\{t,t'\}$ \citep[see, e.g.,][]{ro83}.
For completeness, a formal proof of (\ref{proof1}) can be found in Online Appendix \ref{app2}.

Accordingly, the before-after effect  $\gamma^{ba}$ can be calculated as the difference between the average treatment effects on the treated (ATT) before and after the reform. The pre-reform ATT can be formalised as
\begin{align*}
	\gamma^{pre} = E[Y^1_0(m) - Y^0_0(m)|D=1,T=0].
\end{align*}
The expected potential outcome $E[Y^1_0(m)|D=1,T=0]$ is directly observed from the data. $E[Y^0_0(m)|D=1,T=0]$ is the counterfactual expected potential outcome, because $Y^0_0(m)$ is never observed for treated individuals before the reform. In our setting, $\gamma^{pre}$ is the average effect of training participation under the mandatory system in the pre-reform period for unemployed persons who mandatorily participate. The pre-reform ATT is identified from observed data as
\begin{align*}
	\gamma^{pre}\stackrel{A1a,Aa2}{=}E\left[\frac{1}{p_{1,0,m}} G(1,0,m) Y \right] - E
	\left[\frac{p_{1,0,m}(X)}{p_{1,0,m} \cdot p_{0,0,m}(X)}G(0,0,m)Y\right]. 
\end{align*}

The post-reform ATT can be indicated by
\begin{align*}
	\gamma^{post} = E[Y^1_1(v) - Y^0_1(v)|D=1,T=1].
\end{align*}
The expected potential outcome $E[Y^1_1(v)|D=1,T=1]$ is directly observed from the data. $E[Y^0_1(v)|D=1,T=1]$ is a counterfactual expected potential outcome, because $Y^0_1(v)$ is never observed for treated individuals in the post-reform period. Here, the parameter $\gamma^{post}$ is the average effect of participation in the post-reform period for participants under the voucher system. The post-reform ATT is identified from observed data as
\begin{align*}
	\gamma^{post}\stackrel{A1a,A2a}{=}& \quad E\left[\frac{1}{p_{1,1,v}}G(1,1,v)Y \right] - E\left[\frac{p_{1,1,v}(X)}{p_{1,1,v} \cdot p_{0,1,v}(X)}G(0,1,v)Y\right].
\end{align*}

Next, we focus on the selection effect. In our setting, programme participants before and after the reform are likely to differ in their observed characteristics due to changes in the selection criteria. We are interested in the differences between the effectiveness of training that comes solely by the changing characteristics of participants holding everything else constant on the pre-reform situation,
\begin{align*}
	\gamma^{sel}= E[Y^1_0(m)-Y^0_0(m)|D=1,T=1]-E[Y^1_0(m)-Y^0_0(m)|D=1,T=0].
\end{align*}
The expected potential outcome $E[Y^1_0(m)|D=1,T=0]$ is directly observed from the data. The selection effect is identified under Assumption 1a and 2a by%\footnote{In general, it would be also possible to identify $\gamma_{v,1}^{sel}$ under Assumptions 1a and 2a which would give us the selection effect holding everything constant on post-reform conditions. However, this is not of primary interest in our application.}
\begin{align*}
	\gamma^{sel} \stackrel{A1a,A2a}{=} & \qquad E\left[\frac{p_{1,1,v}(X)}{p_{1,1,v} \cdot p_{1,0,m}(X)}G(1,0,m) Y\right] - E\left[\frac{p_{1,1,v}(X)}{p_{1,1,v} \cdot p_{0,0,m}(X)}G(0,0,m)Y\right]\\
	&  \qquad - \left[E\left[\frac{1}{p_{1,0,m}}G(1,0,m)Y\right] - E\left[\frac{p_{1,0,m}(X)}{p_{1,0,m} \cdot p_{0,0,m}(X)}G(0,0,m)Y\right]\right].
\end{align*}

The identification of business cycle effects and the policy effect requires two additional assumptions because we never observe the pre-reform policy system after the reform and the post-reform policy system before the reform. First, we assume that potential outcomes of the non-treated are independent of the policy system, i.e., we assume that the reform has no effects on the outcomes of the untreated. This is a plausible assumption if only a relatively small fraction of the population is affected by the policy system such that general equilibrium effects can be neglected.\footnote{A possible extension is to focus on bounds instead of point-identification \citep[see discussion in, e.g.,][]{kik17,twi17}.}

\bigskip
\noindent \textbf{Assumption 3} \emph{(Exclusion Restriction on Untreated)} %\\
%For all $t \in \{0,1\}$,
\begin{align*}
	& E[Y^0_{1}(v)|D=1, T=1] = E[Y^0_{1}(m)|D=1, T=1].
\end{align*} 

Second, we impose the assumption of common trends. Thereby, we assume the business cycle effects to be independent of the treatment status, i.e., in absence of the reform the time trends of the potential outcomes would be similar under treatment and non-treatment in the mandatory system when the characteristics of the participants would be fixed.

\bigskip
\noindent \textbf{Assumption 4} \emph{(Common Trend Assumption)}.
\begin{equation*}
	\gamma^{bc0}  = \gamma^{bc1}
\end{equation*}

Under Assumptions 1a, 2a, 3, and 4, we can identify the business cycle effect under mandatory treatment $\gamma^{bc1}$ from observed data as,
\begin{align*}
	\gamma^{bc1} \stackrel{A4}{=} \gamma^{bc0} &=  E[Y^0_1(m)-Y^0_0(m)|D=1,T=1] \\
	&\stackrel{A3}{=} E[Y^0_1(v)-Y^0_0(m)|D=1,T=1] \\
	&\stackrel{A1a,A2a}{=}E\left[\frac{p_{1,1,v}(X)}{p_{1,1,v} \cdot p_{0,1,v}(X)}G(0,1,v)Y\right] - E\left[\frac{p_{1,1,v}(X)}{p_{1,1,v} \cdot p_{0,0,m}(X)}G(0,0,m)Y\right].
\end{align*}

Now, we focus on the parameter of primary interest in this study. The policy effect is the difference of potential outcomes of treated due to a change in the policy instrument from a mandatory to a voucher system holding individual characteristics and time constant on the post-reform situation. By adding and subtracting $E[Y^0_1(v)|D=1, T=1]$ and using $E[Y^0_1(v)|D=1, T=1] = E[Y^0_1(m)|D=1, T=1]$ (A3), we can rewrite the policy effect as
\begin{align*}
	\gamma^{p} & =E[Y^1_1(v)-Y^1_1(m)|D=1,T=1]\\
	&\stackrel{A3}{=}E[Y^1_1(v)-Y^0_1(v)|D=1,T=1]-E[Y^1_1(m)-Y^0_1(m)|D=1,T=1].
\end{align*}

The potential outcome $Y^1_1(m)$ is never observed for treated individuals after the reform. However, under the imposed assumptions the policy effect can be decomposed into the different reform parameters by adding and subtracting 
$E\left[Y^1_0(m) - Y^0_0(m)|D=1,T=0\right]$ and $E\left[Y^1_0(m) - Y^0_0(m)|D=1,T=1\right]$. Thus, the policy effect is equal to the overall reform effect minus business cycle effects minus the selection effect, which are all - as shown above - identified from observed data: 
%\footnote{Adding and substracting $E\left[Y^1_1(v) - Y^0_1(v)|D=1,T=1\right]$ and $E\left[Y^1_1(v) - Y^0_1(v)|D=1,T=0\right]$ would give us $\gamma_{1,0}^{in} = \gamma^{ba} + \gamma^{bc}_1 - \gamma^{bc}_0 - \gamma^{sel}_{v,1}$ holding time constant on pre-reform. It follows from Assumptions 1a and 2a that $\gamma^{in}_{0,1} = \gamma^{in}_{0,0} = 0$.}
\begin{align*}
	\gamma^{p}= & \quad E[Y^1_1(v)-Y^0_1(v)|D=1,T=1]-E[Y^1_1(m)-Y^0_1(m)|D=1,T=1]\\
	& \quad + E\left[Y^1_0(m) - Y^0_0(m)|D=1,T=1\right] - E\left[Y^1_0(m) - Y^0_0(m)|D=1,T=1\right]\\
	& \quad + E\left[Y^1_0(m) - Y^0_0(m)|D=1,T=0\right] - E\left[Y^1_0(m) - Y^0_0(m)|D=1,T=0\right] \\[1em] 
	= & \quad \underbrace{E\left[Y^1_1(v) - Y^0_1(v)|D=1,T=1\right] - E\left[Y^1_0(m) - Y^0_0(m)|D=1,T=0\right]}_{\gamma^{ba}}\\
	& \quad - \underbrace{\Big[E\left[Y^1_0(m)-Y^0_0(m)|D=1,T=1\right]- E\left[Y^1_0(m)-Y^0_0(m)|D=1,T=0\right]\Big]}_{\gamma^{sel}}\\
	&\quad-\underbrace{E\left[Y^1_1(m)-Y^1_0(m)|D=1,T=1\right]}_{\gamma^{bc1}}+\underbrace{E\left[Y^0_1(m)- Y^0_0(m)|D=1,T=1\right]}_{\gamma^{bc0}}.
\end{align*}
Accordingly, $\gamma^{p} = \gamma^{ba}  - \gamma^{sel} - (\gamma^{bc1} - \gamma^{bc0})$, which is the additive separability assumption imposed in \cite{ri13}. We show the sufficient conditions to achieve additive separability. Imposing assumptions 1a, 2a, 3 and 4, we have shown that the total change in the effectiveness of the policy instrument from before to after the reform can be decomposed into the effect of changing the selection, a time effect and the policy effect and that these, in turn, are identified from observed data. Thus, the policy effect can be estimated from observed data as
\begin{align*}
	\gamma^{p}\stackrel{A1a,A2a,A3,A4}{=} & \quad E\left[\frac{1}{p_{1,1,v}}G(1,1,v)Y \right] - E\left[\frac{p_{1,1,v}(X)}{p_{1,1,v} \cdot p_{0,1,v}(X)}G(0,1,v)Y\right]\\
	- & \quad E\left[\frac{p_{1,1,v}(X)}{p_{1,1,v} \cdot p_{1,0,m}(X)}G(1,0,m) Y\right] - E\left[\frac{p_{1,1,v}(X)}{p_{1,1,v} \cdot p_{0,0,m}(X)}G(0,0,m)Y\right].\\
\end{align*}

\subsection{Identification of causal channels}\label{sec23:channels}
We apply a mediation framework \citep[see, for instance, the seminal paper by][]{bar86} to isolate the causal channels through which the policy effect works. In our setting, we aim to separate the effects of voluntary participation (in the following 'assignment effect') from the effect of increased course choice (in the following 'composition effect'). Thereby, we consider the type and duration of training as so-called mediators, i.e., intermediate outcomes on the causal path of the assignment system to the individual labour market outcomes. Let $C$ denote the composition of programmes. To investigate the reform channels, we augment the notation of the potential outcomes with programme composition. This new notation of potential outcomes is directly linked to the former notation by $Y^d_t(s) = Y^d_t(s,C=c)= Y^d_t(s,c)$. We start with the policy effect expressed as the total effect of the change from a mandatory to a voucher system by
\begin{equation}\label{equ2}
\gamma^{p} = E\left[Y^1_1(v,c_v) - Y^1_1(m,c_m)|D=1,T=1\right],
\end{equation}
where we denote $c_m$ the realised programme composition under mandatory assignment and $c_v$ the realised programme composition under voucher assignment.

This extended notation allows us to define further parameters of interest. The impact of the policy effect may be (partly) due to increased course choice or to a direct effect of voluntary participation. In the following, we show how these two effects can be disentangled. First, we are particularly interested in the so-called controlled direct effect \citep[see, for instance,][]{pea01}.  It can be formalised as
\begin{align*}
	\rho  = & \quad E\left[Y^1_1(v,c_v) - Y^1_1(m,c_v)|D=1,T=1\right].
\end{align*}
This is the direct effect of the voucher system for the type and duration composition of training as under the mandatory system, i.e., the assignment effect. Second, the effect of increased course choice can be formalised as
\begin{align*}
	\delta  = & \quad E\left[Y^1_1(m,c_v) - Y^1_1(m,c_m)|D=1,T=1\right].
\end{align*}
This is the indirect effect of increased course choice, i.e., the assignment system is kept constant while the composition of programme types and durations varies. As can be seen from adding and substracting $Y^1_1(m,c_v)$ in the expectation of expression (\ref{equ2}), the direct effect $\rho$ and the indirect effect $\delta$ sum up to the total policy effect $\gamma^{p}$.

However, causal mechanisms are not easily identified. Even if the policy effect is identified, this would not imply identification of the mediator effects. Addressing the endogeneity of mediators requires that they are independent of the potential outcomes conditional on the policy system and the covariates. %In our application, this assumption implies that training participants with more favourable unobserved labour market opportunities do not systematically select into programmes of a specific type or duration.  

\bigskip
\noindent \textbf{Assumption 1b} \emph{(Sequential Conditional Mean Independence)}\\
For all $s,s'\in \{m,v\}$ and $x \in \mathbb{X}$,
\begin{align*}
	& E[Y^1_1(s,c_{s'})|D=1,T=1,C=c_{s'},X=x] = E[Y^1_1(s,c_{s'})|D=1,T=1,X=x]
\end{align*}
and all necessary moments exist.
\bigskip 

Assumption 1b implies for treated in the post-reform period that, given the observed pre-treatment confounders, the expected potential outcomes are independent of the type and duration of training. The selection of the type and duration of training causally succeeds the selection into treatment. Therefore, we call Assumption 1b sequential conditional mean independence. The combination of Assumptions 1a and 1b is analogue to sequential conditional independence assumption invoked in the non-parametric mediation literature for identifying direct effects \citep[see, e.g.,][]{ima11}. In contrast, a multiple treatment framework would assume contemporaneous selection into treatment and selection of the type and duration of training \citep[][]{imb00,le01}. Then Assumptions 1a and 1b would have to hold contemporaneously instead of sequentially.  

\noindent \textbf{Assumption 2b} \emph{(Support)}.
\begin{equation*}
	0< Pr(G(d,t,s)=1|C=c,X=x) <1 \ \  \forall \ \  s\in\{m,v\}, d,t\in \{0,1\}.
\end{equation*}

Assumption 2b requires overlap in the propensity score distributions of the mediators under both systems and control variables. Finally, under Assumption 1a,b, 2a,b, 3 and 4 the controlled direct and the indirect effects can be identified as 
\begin{align*}
	\rho \stackrel{A1a,b,A2a,b,A3,A4}{=}& \quad E\left[\frac{1}{ p_{1,1,v}}G(1,1,v)Y \right] - E\left[\frac{p_{1,1,v}(X)}{p_{1,1,v} \cdot p_{0,1,v}(X)}G(0,1,v)Y\right]\\
	- & E\left[\frac{p_{1,1,v}(X,C)}{p_{1,1,v} (C)\cdot p_{1,0,m}(X,C)}G(1,0,m) Y\right] - E\left[\frac{p_{1,1,v}(X)}{p_{1,1,v}\cdot p_{0,0,m}(X)}G(0,0,m)Y\right],\intertext{and}
	\delta \stackrel{A1a,b,A2a,b,A3,A4}{=} & \quad  \gamma^{p} - \rho,
\end{align*}
with $p_{1,1,v}(x,c)= Pr(G_i(1,1,v)=1|C=c,X_i=x)$ and $p_{1,1,v}(c)= Pr(G_i(1,1,v)=1|C=c)$ \citep[see, e.g.,][]{hub14}.

\section{The reform of the allocation of vocational training}\label{sec3:training}

We illustrate our approach using a large-scale reform of the allocation system of unemployed individuals to vocational training in Germany. This reform presents an illustrative example in which policy effects, selection effects, and time effects are part of the overall reform effect. The main objective of vocational training for unemployed persons is the adjustment of their skills to changing requirements in the labour market and/or changed individual conditions (due to health problems, for example). In Germany, vocational training comprises three types of programmes: practice firm training, classical vocational training, and retraining. Classical vocational training courses takes place in classrooms or on-the-job and are categorised by their planned durations. We distinguish between short training (a maximum duration of six months) and long training (a minimum duration of six months). Practice firm training simulates a work environment in a practice firm. Retraining (also called degree course) has long durations of up to three years. It leads to the completion of a (new) vocational degree within the German apprenticeship system. Further descriptions and examples of courses can be found in Table \ref{programs}.

\[ \text{Table \ref{programs} around here} \]

\subsection{The reform} \label{sec31:reform}
%Before 2003, strong authority and control of caseworkers regarding the participation decision and the choice of training providers and courses characterised the allocation process into vocational training. Caseworkers' assignment of unemployed to courses was mandatory and based on subjective criteria. Close cooperations between local employment agencies and training providers were strongly critized to result in a supply determined allocation of unemployed into courses.  

Before 2003, caseworkers' assignment of unemployed to courses was mandatory and based on subjective criteria.
The introduction of a voucher system on January 1, 2003 had the intention to increase the responsibility of training participants and to establish market systems for training providers \citep{bru05}. Potential training participants receive a training voucher that allows them to select the provider and course. Their choice is subject to the following restrictions: First, the voucher specifies the objective, content, and maximum duration of the course. Second, it can be redeemed within a one-day commuting zone. Third, the validity of training vouchers varies between one week and a maximum of three months. Importantly, caseworkers cannot impose sanctions if a voucher is not redeemed.

Simultaneously with the voucher system, the reform introduced stricter selection criteria for potential training participants. The post-reform paradigm of the German Federal Employment Agency focuses on direct and rapid placement of unemployed individuals, high reintegration rates, and low dropout rates. Caseworkers award vouchers such that at least 70\% of all voucher recipients are expected to find jobs within six months of completing training.\footnote{The enforcement of the 70\% criterion was difficult, because satisfying the rule had no consequences. For this reason, the selection rule was abolished after 2004.} 

\subsection{Data, treatment and sample} \label{sec32:data}
This study is based on administrative data provided by the German Federal Employment Agency. The data set contains information on \emph{all} individuals in Germany who participated in a training programme between 2001 and 2004. Individual records are collected from the Integrated Employment Biographies (IEB).\footnote{The IEB is a rich administrative database and the source of the sub-samples of data used in all recent studies that evaluate German ALMP programmes \citep[e.g.,][]{bi12,le11,le13}. The IEB is a merged data file containing individual records collected from four different administrative processes: the IAB Employment History ({\it Besch\"aftigten-Historik}), the IAB Benefit Recipient History ({\it Leistungsempf\"anger-Historik}), the Data on Job Search originating from the Applicants Pool Database ({\it Bewerberangebot}), and the Participants-in-Measures Data ({\it Ma\ss{}nahme-Teilnehmer-Gesamtdatenbank)}. IAB ({\it Institut f\"ur Arbeitsmarkt- und Berufsforschung}) is the abbreviation for the research department of the German Federal Employment Agency.} The sample used as the comparison group originates from the same database. It is constructed as a 3\% random sample of individuals who experience at least one transition from employment to non-employment.\footnote{We account for the fact that we have different sampling probabilities in all calculations whenever necessary.} 

\begin{figure}[h!]
\begin{center}
\caption{Graphical illustration of Sample A and B} \label{samples}
\renewcommand{\baselinestretch}{1}
\begin{tabular}{c}\subfigure[Sample A]
{\includegraphics[width=12cm]{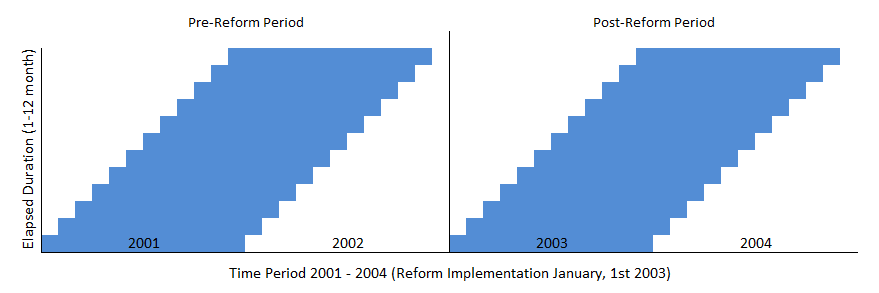}}\\
\subfigure[Sample B]
{\includegraphics[width=12cm]{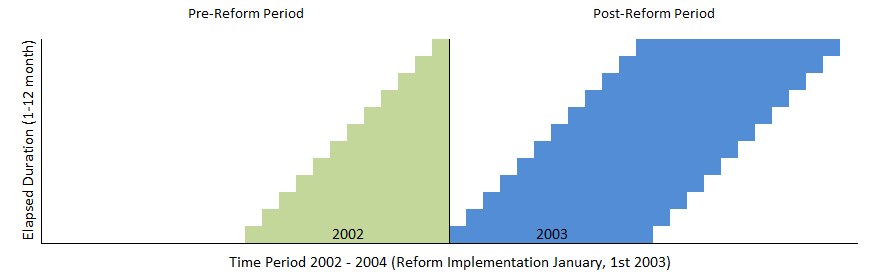}}
\end{tabular}
\end{center}
\end{figure}

The treatment is defined as the first participation in a vocational training programme during the first year of unemployment. We follow a static evaluation approach and impute (pseudo) participation starts \citep[similar to, e.g.,][]{le99,le07}. The evaluation sample is constructed as an inflow sample into unemployment. The baseline sample (Sample A) consists of individuals who became unemployed in 2001 under the mandatory system or in 2003 under the voucher system, after having been continuously employed for at least three months. Additionally, we use an alternative sample definition (Sample B) for which we alter the pre-reform sample restrictions. We consider individuals who enter unemployment in 2002 and start training within the following 12 months but no later than December 2002. Thereby, we approximate the timing of the reform implementation with respect to inflow into unemployment. Sample B is used for robustness tests. A graphical illustration of the samples is presented in Figure \ref{samples}.

\[ \text{Figure \ref{samples} around here} \]

Entering unemployment is defined as the transition from (non-subsidised, non-marginal, non-seasonal) employment to non-employment of at least one month. We focus on individuals who are eligible for unemployment benefits at the time of inflow into unemployment. This sample choice reflects the main target group of vocational training. We only consider individuals aged between 25 and 54 years at the beginning of their unemployment spell to exclude individuals who are eligible for specific labour market programmes targeting youths and individuals eligible for early retirement schemes.

\subsection{Descriptive statistics}\label{sec33:descriptives}
The baseline Sample A includes 206,511 unweighted or 1,011,125 reweighted observations. We account for the fact that we use a 100\% sample of programme participants and a 3\% random sample of non-participants using the inverse inclusion probabilities as weights. We observe 26,341 unemployed individuals who redeem vouchers and 69,216 participants who are directly assigned to a training course. This is the full sample of vocational training participants in Germany that satisfies our sample selection criteria. The sample includes 420,014 reweighted control persons before and 495,554 reweighted control persons after the reform. 

\[ \text{Table \ref{descriptives} around here} \]

In Table \ref{descriptives}, we report the sample first moments of the observed characteristics with a large standardised difference. Additionally, we present descriptive statistics for observed characteristics with small standardised differences in Table \ref{descriptives_app} in Online Appendix \ref{app1}.
In the first two columns of Table \ref{descriptives}, we report the sample first moments of the control variables for participants and non-participants under the voucher system. The respective sample moments under the mandatory system can be found in the third and fourth columns. The last three columns display the standardised differences between the different sub-samples and the treatment group under the voucher system. Training participants are on average younger, have fewer instances of incapacity and are better educated. They have more successful employment and welfare histories than unemployed individuals in the comparison group. These patterns are observed under both systems. The primary differences are observed in the employment histories of participants and the regional characteristics. Training participants under the voucher system have been employed longer and have higher cumulative earnings than participants under the mandatory system. Furthermore, participants under the voucher system are more likely to reside in local employment agency districts with low employment in the construction sector and a high share of male unemployment.

\subsection{Plausibility of identifying assumptions}\label{sec34:assumptions}
Assumptions 1a, 1b are strong, but standard in the programme evaluation literature. The plausibility of similar assumptions has been studied by \nocite{bi12}Biewen et al. (2014) and \cite{le13} for training programme evaluations. Their findings suggest that such assumptions are plausible for training programme evaluations when rich data is available. We use exceptionally rich data, which includes the control variables used in the previous literature and additional new variables. In particular, we use baseline personal characteristics, the timing of programme starts, regions, benefit and unemployment insurance claims, pre-programme outcomes, and labour market histories (see Table \ref{descriptives} and Table \ref{descriptives_app} in Online Appendix \ref{app1}). In addition to the standard variables, we control for proxy information concerning physical or mental health problems, lack of motivation, and reported sanctions. Furthermore, we control for regional characteristics at the level of local employment agency districts, which are often not available with such precision. Thus, the imposed assumptions appear to be plausible in our setting.

%Regarding Assumptions 1a,b, we argue that they are more plausible the shorter the time difference between the before and the after-reform period is and the better we control for economic conditions. Therefore, in our main specifications, we control for characteristics of local employment agency districts at treatment start as a sensitivity test for this assumption. Moreover, we use samples with different calendar periods as robustness checks (see Section \ref{sec32:data}). 

Assumption 2a,b can be tested using the data. In unreported calculations, we perform simple support tests and do not observe any incidence of support problems. 

Assumption 3 requires that the reform has no effect on the non-treated. After controlling for the changed selection of treated before and after the reform, which can indeed change the composition of the non-treated, it is plausible that the assignment system is independent of the potential outcomes of non-participants. The main argument for this is that the reform of the assignment mechanism only affects participants and the share of participants is relatively small, such that general equilibrium effects can be neglected.

\begin{figure}[h!]
\begin{center}
\setcounter{subfigure}{0}\caption{Time trends of employment probabilities for different subgroups of individuals for the 1991-2012 period}\label{timetrend1}
\renewcommand{\baselinestretch}{1}
\begin{tabular}{c}
\includegraphics[width=10cm]{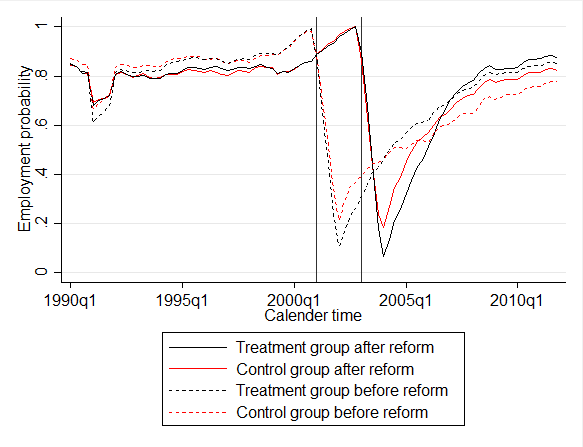} \\
\end{tabular}
\parbox{16cm}{\scriptsize Note: We report time trends for the years between 1990 and 2012. The outcome variables are reweighted as described in Online Appendix \ref{app3}. Similar findings are obtained without reweighting.}
\end{center}
\end{figure}

We show several plausibility tests for Assumption 4, which requires that the potential outcomes of participants and non-participants would follow the same trend in the absence of the reform. % Without this assumption, the policy effect is not identified. 
We present three different types of supporting evidence for the plausibility of this assumption. First, Figure \ref{timetrend1} reports the long-term trends in the outcome variables for different samples for the years between 1990 and 2012. Prior to the treatment start dates in 2001 and 2003, the outcomes of the participants and non-participants samples evolve in parallel over many years. Given these parallel trends, it is likely that we would observe the same respective patterns after 2001 or 2003 in the absence of a treatment.

\[ \text{Figure \ref{timetrend1} around here} \]

Second, we experiment with additional information on local employment agency districts (i.e., regional control variables). We observe the monthly regional unemployment rate (by gender and citizen status), the ratio of vacant full-time jobs, employment shares by sector and population density. We assess the sensitivity of our findings with respect to these factors.
% In the specifications with regional controls, we adjust for possible differences in the regional unemployment rates between the pre- and post-reform periods.
If our results are not sensitive to the regional control variables, we expect that possible interactions between the effectiveness of training participation and the unemployment rate (or the business cycle in general) are not important in our application. This would support the plausibility of the common trend assumption.

Third, we use an alternative sample definition (Sample B) for which we alter the pre-reform sample restrictions. We consider individuals who enter unemployment in 2002 and start training within the following twelve months but no later than December 2002. Consequently, not all individuals in Sample B can participate during the first twelve months of their unemployment period (e.g., an individual who enters unemployment in October can only receive treatment under the mandatory system in the following three months). %The post-reform evaluation sample is not altered in Sample B to ensure that the comparison of results for the different samples is straightforward. 
Using Sample B, we approximate the timing of the reform implementation with respect to the inflow into unemployment. We argue that the common trend assumption is more likely to hold if the time difference between the pre- and post-reform periods is smaller. However, in contrast to the baseline sample (Sample A), Sample B is not balanced in the pre- and post-reform periods (comp. Figure \ref{samples}).

\subsection{Estimation}\label{sec35:data}
We apply a semi-parametric reweighting estimator, \emph{Auxiliary-to-Study Tilting} \citep{gr16}, in all estimations. This estimator is well suited to our empirical design because it balances the efficient sample first moments exactly. Furthermore, it is $\sqrt{N}$-consistent and asymptotically normal. The estimator is described in Online Appendix \ref{app3}.

\subsection{Empirical results}\label{sec36:results}

\subsubsection{Decomposition of reform effect into selection, time and policy effects}\label{sec361:policy}

\begin{figure}[h!]
\begin{center}
\caption{Post-reform, pre-reform ATTs and overall reform effect} \label{reform1_emp}
\renewcommand{\baselinestretch}{1}
\begin{tabular}{c}
\includegraphics[width=10cm]{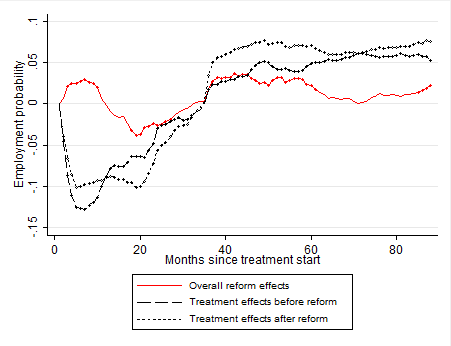} \\
\end{tabular}
\parbox{16cm}{\scriptsize Note: We estimate separate effects for each of the 88 months following the treatment. Diamonds indicate significant point estimates at the 5\%-level. Significance levels are bootstrapped with 499 replications. Lines without diamonds indicate point estimates that are not significantly different from zero. We use baseline Sample A and control for local employment agency district characteristics and the full set of observed characteristics (see Table \ref{balance} in Online Appendix \ref{app1}).}
\end{center}
\end{figure}

We start this section by showing the overall reform effect. Figure \ref{reform1_emp} presents the ATTs for participants in vocational training courses before the reform ($\gamma^{pre}$) and after the reform ($\gamma^{post}$). The outcomes of interest are nonsubsidised and nonmarginal employment which is subject to social security contributions (`employment' in the following). Results for monthly earnings are available in the online appendix.\footnote{Subsidized employment is employment in the context of an ALMP. Marginal employment is according social security regulations in Germany defined as employment of a few hours per week only.} We report separate effects for every month during 88 months following the course start. The lines are monthly point estimates and the diamonds indicate significant effects at the 5\% level.

\[ \text{Figure \ref{reform1_emp} around here} \]

Training participants suffer from negative lock-in effects before and after the reform. The lock-in effects are steeper in the pre-reform period but have longer durations after the reform. The long-term effects of participation in vocational training courses on employment probability are positive. Training participation increases long-term employment probability (seven years after the start of training) by five percentage points before the reform and by 7.5 percentage points after the reform. 

The raw difference between the post- and pre-reform effectiveness of training identifies the overall difference in effects before and after the reform ($\gamma^{ba}$). In Figure \ref{reform1_emp}, the red solid line shows a positive difference in effects before and after the reform in the short-term and negative effects in the second and third years after the course start. In the long-term (seven years after the course start), the difference between the post- and pre-reform effectiveness of training is significant and positive. The reform increased the employment probability by 2-3 percentage points seven years after vocational training participation starts. This overall difference is the starting point of our analysis and will be decomposed into the individual effects of stricter participant selection, business cycle effects and the policy effects of the changing allocation system from a mandatory to a voucher system.

%Even though these effects seem small, they reflect 25-60\% of the main effects of vocational training (the main effects show between 5-7.5 percentage points higher employment chances). 

\begin{figure}[h!]
\begin{center}
\caption{Selection and overall reform effects}\label{reform2_emp}
\renewcommand{\baselinestretch}{1}
\begin{tabular}{c}
\includegraphics[width=10cm]{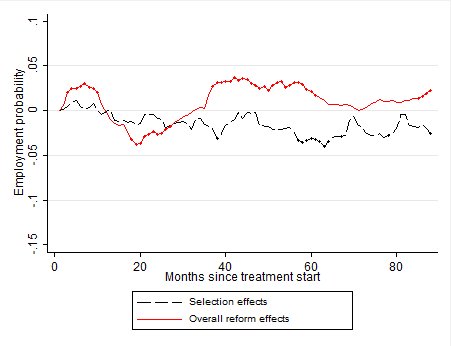} \\
\end{tabular}
\parbox{16cm}{\scriptsize Note: We estimate separate effects for each of the 88 months following the treatment. Diamonds indicate significant point estimates at the 5\%-level. Significance levels are bootstrapped with 499 replications. Lines without diamonds indicate point estimates that are not significantly different from zero. We use baseline Sample A and control for local employment agency district characteristics and the full set of observed characteristics (see Table \ref{balance} in Online Appendix \ref{app1}).}
\end{center}
\end{figure}

%\subsubsection{Selection effects}
The imposition of stricter selection criteria changes the composition of training participants with respect to their labour market characteristics. Because caseworkers are instructed to assign training to unemployed individuals with high re-employment probabilities, we expect to observe training participants with better labour market characteristics after the reform. In Table \ref{balance} in Online Appendix \ref{app1}, we report the efficient first moments of all confounding control variables for training participants before and after the reform. The largest differences between the two groups can be found for the employment and welfare histories and the characteristics of local employment agency districts. Unemployed persons who participate in the voucher system, i.e., after the reform, have on average more successful employment and earnings profiles than those who participated in the mandatory system. 

\[ \text{Figure \ref{reform2_emp} around here} \]

The impact of stricter selection criteria on the effectiveness of training can be captured by the selection effects ($\gamma^{sel}$), which are reported in Figure \ref{reform2_emp}. The effects show the differences in the effectiveness of training that can be solely explained by a different participant selection in terms of their characteristics holding time and policy instrument constant. The results suggest that stricter selection criteria only have a minor influence on the effectiveness of training. If anything, we find negative selection effects over the long-term. Given the small differences in most observed characteristics, such small and mostly insignificant selection effects are plausible.

\begin{figure}[h!]
\begin{center}
\caption{Time effects}
\label{reform3_emp}
\renewcommand{\baselinestretch}{1}
\begin{tabular}{c}
\includegraphics[width=10cm]{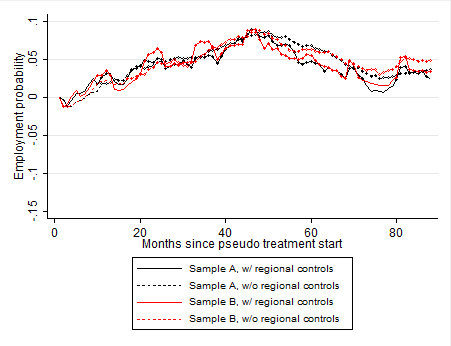} \\
\end{tabular}
\parbox{16cm}{\scriptsize Note: We estimate separate effects for each of the 88 months following the treatment. Diamonds indicate significant point estimates at the 5\%-level. Significance levels are bootstrapped with 499 replications. Lines without diamonds indicate point estimates that are not significantly different from zero. We use baseline Sample A and control for local employment agency district characteristics and the full set of observed characteristics (see Table \ref{balance} in Online Appendix \ref{app1}).}
\end{center}
\end{figure}

Figure \ref{reform3_emp} presents the business cycle effects of non-participation ($\gamma^{bc0}$) for Samples A and B with and without additional regional control variables. The time effects show an immediate, sharp increase of employment probabilities which peaks after three years. Thereafter, the effects evolve to a 3-5 percentage points higher employment probability in the post-reform period compared to the pre-reform period. 

%The time effects for monthly earnings evolve more smoothly over the observation period. After seven years, the difference in monthly earnings between the post- and pre-reform periods amounts to 50-100 Euros.%\footnote{These findings also support the plausibility of conditional mean independence and the exclusion restriction. Potential outcomes under non-treatment are significantly different between the two time periods only after more than one year (excluding specifications without regional labour market characteristics). This suggests that there are no systematic differences of non-treated at (pseudo) treatment start or shortly thereafter. Long-term differences in the economic conditions do not violate the conditional mean independence assumption, as these differences have no influence on the treatment probability (if they could not be anticipated).} 

\[ \text{Figure \ref{reform3_emp} around here} \]

The general pattern of the time effects is not sensitive to the sample definition or to the inclusion of additional regional labour market characteristics. This supports the plausibility of the common trend assumption. However, by the implementation of the Hartz reforms, the German labour market was intensively reformed during the observation period, particularly in 2005. An improvement of labour market conditions can be observed over the long-term. This does not alter the plausibility of our identifying assumptions as long as all groups are equally affected by the Hartz reforms.

\begin{figure}[h!]
\begin{center}
\caption{Policy effects}
\label{reform5_emp}
\renewcommand{\baselinestretch}{1}
\begin{tabular}{c}
\includegraphics[width=10cm]{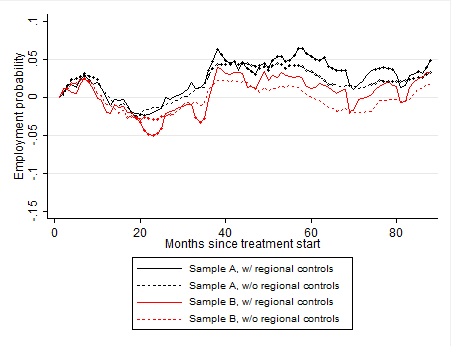} \\
\end{tabular}
\parbox{16cm}{\scriptsize Note: We estimate separate effects for each of the 88 months following the treatment. Diamonds indicate significant point estimates at the 5\%-level. Significance levels are bootstrapped with 499 replications. Lines without diamonds indicate point estimates that are not significantly different from zero. We use baseline Sample A and control for local employment agency district characteristics and the full set of observed characteristics (see Table \ref{balance} in Online Appendix \ref{app1}).}
\end{center}
\end{figure}

%\subsubsection{Policy effects}
Finally, Figure \ref{reform5_emp} displays the policy effects for Samples A and B, with and without additional regional control variables. They show the difference in the effectiveness of training that can be solely explained by the changing assignment system from mandatory assignment to vouchers holding participants characteristics and time period fixed.

%The following thought experiment clarifies the interpretation of the policy effects: Compare the employment outcomes of training participants who receive a training voucher with the employment outcomes that they would obtain if they were mandatory assigned to a (potentially different) vocational training course. Individual labour market characteristics and the time period are fixed.

The pattern of the policy effects varies in the different periods after course start. In the short term, the policy effects are positive, implying that training is more effective under the voucher system. In the best case, training participants who receive a voucher have employment probabilities that are approximately 2-3 percentage points higher compared to participants in the mandatory system. Over the medium term, the policy effects are negative. The specifications using Sample B present a slightly more negative picture. In the worst-case scenario, the employment probability decreases by 5 percentage points. Three years after the start of training, we observe an increase to slightly positive but mostly insignificant policy effects. After seven years, the effects are positive for all specifications. However, the effects are only significant for Sample A  with a 4-5 percentage point increase in employment probabilities.\footnote{The results of all specifications are relatively stable between 40 and 80 months after training participation begins. This mitigates concerns that our findings are greatly altered by the financial crisis in 2008.}

\[ \text{Figure \ref{reform5_emp} around here} \]

\subsection{Channels of the policy effects}\label{sec362:channels}
To interpret the policy effect, it is necessary to investigate the channels through which the reform affect the employment outcome. First, training is voluntary after the reform. Thus, the effectivness of training might differ between a voucher and a mandatory system because voluntarily assigned participants are more motivated than compulsory assigned participants. Second, voucher assigned participants have free course choice conditional on the specification on the voucher. In Table \ref{types}, we report descriptive statistics for different types and duration of training programmes before and after the reform. The share of short training programmes increases from 21\% to 42\% after the reform. Moreover, the share of long training programmes decreases from 41\% to 19\%. The average planned and actual duration of long programmes (practice firm) decrease nearly three (two) months after the reform. The share of participants in retraining courses increases from 19\% to 25\%. The average planned duration is extended by more than one month.\footnote{In 2003, there was also a reduction in the total number of vocational training programmes for political reasons.}

\[ \text{Table \ref{types} around here} \]

Accordingly, the composition of programme types and durations changed substantially after the reform. We observe higher shares of participants in programmes with a duration of less than six months and higher shares of participants in very long programmes with durations of more than two years. The first development might reflect increased freedom of choice under the voucher system. Training vouchers are determined with respect to the maximum programme duration. The unemployed individuals are free to choose a training provider and may self-select into shorter courses. 

%However, after the reform, caseworkers may have an incentive to strategically assign the maximum programme duration to comply with the stricter selection criteria (see the discussion in Section \ref{sec52}). In particular, they have incentives to award vouchers with short maximum duration to unemployed individuals with good employment opportunities. 

To disentangle the effects of voluntary participation from the effects of increased course choice we apply a mediation framework \citep[see, e.g,][]{rob92,bar86}. We consider the type and duration of training as mediators, i.e., intermediate outcomes on the causal path of the allocation system to the individual labour market outcomes. We are particularly interested in the controlled direct effect \citep[see, for instance,][]{pea01}, which is the direct effect of the voucher system for a fixed type and duration of training, i.e., the effect of voluntary participation. It reflects the impact of changing the allocation from a mandatory to a voluntary system if the course composition is held constant for both periods. The indirect effect reveals the isolated impact of the changed composition of programme types and duration after the reform, i.e., the effect of increased course choice. For the estimation of the controlled direct effect, we manipulate the programme durations such that they are similar in the samples of treated individuals before and after the reform by using additional moment conditions for the estimation of the conditional treatment probabilities $p_{1,1,v}(x,c_v)$.\footnote{We generate dummies for the planned programme durations (less than 6 months, between 6 and 12 months, between 12 and 24 months, and more than 24 months). These durations correspond to different programme types. Furthermore, we account for interactions between these dummies and the planned programme duration to allow for linear trends within each period.} 

\begin{figure}[h!]
\begin{center}
\caption{Decomposition of policy effect into course composition (indirect) and voluntary participation (direct) effects} \label{out2_emp}
\renewcommand{\baselinestretch}{1}
\begin{tabular}{cc}
\subfigure[Sample A]{\includegraphics[width=7.5cm]{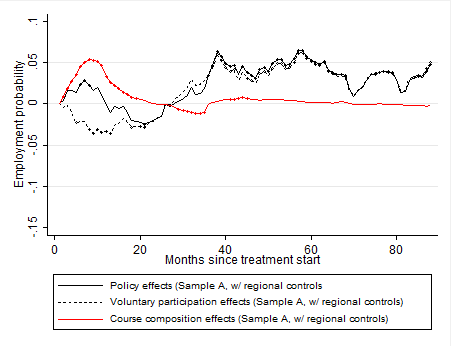}} & 
\subfigure[Sample B]{\includegraphics[width=7.5cm]{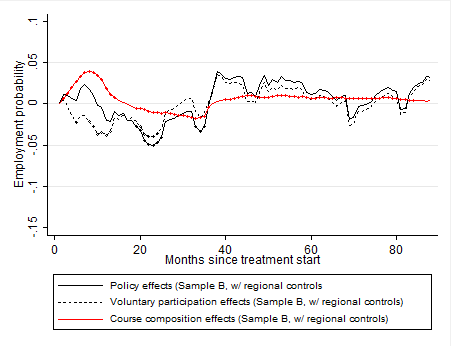}} \\
\end{tabular}
\parbox{16cm}{\scriptsize Note: We estimate separate effects for each of the 88 months following the treatment. Diamonds indicate significant point estimates at the 5\%-level. Significance levels are bootstrapped with 499 replications. Lines without diamonds indicate point estimates that are not significantly different from zero. We use baseline Sample A and control for local employment agency district characteristics and the full set of observed characteristics (see Table \ref{balance} in Online Appendix \ref{app1}). In the duration effects, we account for the planned course durations and interactions using fixed duration dummies.}
\end{center}
\end{figure}

Figure \ref{out2_emp} shows the policy effects, the course composition effects and the effects of voluntary participation for Samples A and B with regional control variables. We find positive short-term effects that can be explained by the larger share of short programmes after the reform. After 2-3 years, the effects turn negative which can be explained by a larger share of retraining programmes in the voucher system. In the long term, the course composition effects become slightly positive but remain close to zero.

\[ \text{Figure \ref{out2_emp} around here} \]

The effects of voluntary participation become negative immediately after the start of training. After two years, voluntary participation leads to a 3-5 percentage points decline in the employment probability compared to mandatory participation. The voluntary participation effects remain negative until three years after the start of training. Unemployed individuals might perceive less pressure to find a job under voluntary participation, as they feel more accommodated, have more positive attitudes towards the training course and a higher motivation to complete the programme. A descriptive analysis of dropout rates supports this interpretation (see Online Appendix \ref{app_dropout}). We find that the dropout rates decrease by more than four percentage points after the reform. This may negatively affect job search intensity, which could lead to more pronounced negative lock-in effects and may raise participants' reservation wages. Job search intensity and reservation wages have opposite effects on realised earnings. 

%\footnote{It is not possible to include the dropout rate as additional outcome variable in our empirical design because course completion is only observed for participants. Therefore, we perform a probit estimation of the dropout rate on a post-reform dummy and all other control variables. We use the dropout definition proposed by \cite{paul15} and count an observation as dropout if less than 80\% of the planned programme duration is completed. The regression results are presented in Table \ref{dropout} in Online Appendix \ref{app_dropout}.} 
%Furthermore, assignment to tedious training programmes and sanction possibilities are limited under voluntary participation.

Three years after programme entry, the voluntary participation effects become positive. Using Sample A, we report a five percentage point increase in employment probabilities. These findings are stable over the considered time period in this study. In the more conservative specification (Sample B), we do not observe any significant long-term impacts. However, the patterns tend to be positive. This suggests that participants accumulate more human capital under voluntary participation, and the pay-offs of these investments need time to unfold.

%Finally, we consider effect heterogeneity by programme type in Figure \ref{out5} and \ref{out6} in Online Appendix \ref{app5}. Overall, the employment probabilities and monthly earnings are lower for participants in longer programmes (long training and retraining) than for those in shorter programmes (practice firm training and short training). It appears that the voluntary participation is more effective for shorter than for longer vocational training programmes in the time period under investigation. The effects of voluntary participation increases significantly the employment probabilities of participants in parctice firm training and short training programmes. 

\subsection{Discussion}\label{sec37:discussion}
Our results qualitatively confirm the findings in \cite{ri13} for the time horizon of 1.5 years after treatment. We find positive effects of the reform of the allocation system in the short-term. Moreover, we find that the reform of the allocation system reduces the re-employment probabilities between the first and second year after the start of training. Our application shows that the consideration of long-term effects is crucial. In the long-term, the policy effects turn positive and remain on an approximately stable level until seven years after the training started.

Compared to earlier studies, we show that it is important to consider direct and indirect effects of a policy reform. We provide evidence that the short-term positive policy effects are mainly driven by a changing composition of training course types and duration after the reform. The share of individuals who participate in shorter courses increased in the post-reform period. The selection into shorter courses improves the labour market outcomes in the short-term. This is almost a mechanical effect, because participants in shorter courses are distracted from intensive job search for a shorter time period. 

If we focus on the direct effect of voluntary participation net of the course composition effect, we observe a reduction of training effectiveness in the short-term and a significant increase in the long run. This can be explained by a higher motivation of participants under the voucher system to focus on the course contents and to complete training instead of intensively search for a new job during course participation.

\section{Conclusion} \label{sec4:conclusion}
In this study, we formally show the identification of channels of policy reforms with multiple treatments and different selection into each type of treatment. We discuss the assumptions that are sufficient to identify the different components of the policy reform which are selection effects, time effects and the policy effects. Furthermore, we provide a formal framework of the causal channels through which the policy effects may affect the outcome of interest using mediation analysis. 

We illustrate the empirical approach using a large reform of the allocation of vocational training programmes in Germany. The pre-reform system granted caseworkers substantial authority through mandatory allocation of unemployed individuals to training courses. The post-reform voucher system introduces voluntary participation and some freedom of course choice. Additionally, the reform changed the criteria for selecting unemployed persons into training programmes. This reforms is a illustrative example in which the overall reform effect can be decomposed into selection effects, time effects and the policy effects of interest. We separate the different reform components from each other and investigate the channels through which the reform of the allocation system operates. We are mainly interested in the direct effect of changing the allocation of vocational training from a mandatory to a voluntary system net from indirect effects that may occur through the increased course choice.

The empirical results show the importance of considering causal channels since they may operate in opposing directions. Here, the policy effect indicates an increased effectiveness of training after the reform in the short run. We show that the positive effect mainly comes from indirect effects of the policy reform whereas the direct effects show a short-term reduction in the effectiveness of training. This is important knowledge for policy makers because it allows to target policy instruments more precicely. Depending on the short- and long-term objectives of policy makers it may even reverse the application of policy instruments.

%\bigskip

%\noindent \textbf{Compliance with Ethical Standards:}

%\noindent \textbf{Funding:} This study was funded by the Institute for Employment Research (project 1155 ``Regional Allocation Intensities, Effectiveness and Reform Effects of Training Vouchers in Active Labor Market Policies''). 

%\noindent \textbf{Ethical approval:} This article does not contain any studies with human participants performed by any of the authors.

%%%%%%%%%%%%%%%%%%
%%% REFERENCES %%%
%%%%%%%%%%%%%%%%%%

\bibliographystyle{econometrica}
\bibliography{Bibliothek}
\clearpage

%%%%%%%%%%%%%%%
%%% FIGURES %%%
%%%%%%%%%%%%%%%

\section*{Figures}

%%%%%%%%%%%%%%%%
%%% FIGURE 1 %%%
%%%%%%%%%%%%%%%%

%%%%%%%%%%%%%%%%
%%% FIGURE 2 %%%
%%%%%%%%%%%%%%%%

\setcounter{subfigure}{0}

%%%%%%%%%%%%%%%%
%%% FIGURE 3 %%%
%%%%%%%%%%%%%%%%

%%%%%%%%%%%%%%%%
%%% FIGURE 4 %%%
%%%%%%%%%%%%%%%%

\setcounter{subfigure}{0}

%%%%%%%%%%%%%%%%
%%% FIGURE 5 %%%
%%%%%%%%%%%%%%%%

\setcounter{subfigure}{0}

%%%%%%%%%%%%%%%%
%%% FIGURE 6 %%%
%%%%%%%%%%%%%%%%

\setcounter{subfigure}{0}

\clearpage

%%%%%%%%%%%%%%%
%%% TABLES %%%%
%%%%%%%%%%%%%%%

\section*{Tables}

%%%%%%%%%%%%%%%%%%%%%
%%%%% TABLE 1 %%%%%%%
%%%%%%%%%%%%%%%%%%%%%

\begin{table}[h!]
\centering \caption{Vocational training programmes}
\renewcommand{\arraystretch}{1}
 \label{programs} \scriptsize
\begin{tabularx}{\textwidth}{lXX}
\noalign{\smallskip}\hline\hline\noalign{\smallskip}
&  &\\
Programme type & Description & Examples\\
\noalign{\smallskip}\hline\hline\noalign{\smallskip}
Practice firm training & Courses that took place in practice firms to simulate a work environment. & Training in commercial software, for office clerks, in data processing\\
\noalign{\smallskip}\noalign{\smallskip}
Short training & Provision of occupation specific skills (duration $\leq$ 6 months). & Training courses for medical assistants, office clerks, draftsman, hairdressers, lawyers \\
\noalign{\smallskip}\noalign{\smallskip}
Long training & Provision of occupation specific skills (duration $>$ 6 months). & Training for tax accountants, elderly care nurses, office clerks, physical therapists\\
\noalign{\smallskip}\noalign{\smallskip}
Retraining & Courses to obtain a first/new vocational degree. & Apprenticeship as elderly care nurses, physical therapists, hotel and catering assistants\\
\noalign{\smallskip}\noalign{\smallskip}
Others & e.g., courses for career improvement & \\
\noalign{\smallskip}\hline\hline\noalign{\smallskip}
\end{tabularx}
\parbox{\textwidth}{\scriptsize Note: We use the categorisation of programmes proposed by \cite{le11}. Additionally, we use the information on the training voucher with regard to the contents of the training courses to construct this table. The examples refer to training goals that are often denoted on the training voucher. The category ''Others'' contains different types of training programmes with few participants.}
\end{table}

%%%%%%%%%%%%%%%%%%%%%
%%%%% TABLE 2 %%%%%%%
%%%%%%%%%%%%%%%%%%%%%

\begin{table}[h!]
\begin{center}
\centering \caption{Sample first moments of observed
characteristics with large standardised differences.}
\renewcommand{\arraystretch}{1}
 \label{descriptives} \scriptsize
\begin{tabularx}{\textwidth}{Xccccccc}
\noalign{\smallskip}\hline\hline\noalign{\smallskip}
& \multicolumn{2}{c}{Voucher system} & \multicolumn{2}{c}{Mandatory system}& \multicolumn{3}{c}{Absolute standardised differences between}\\
& Treatment- & Control- & Treatment- & Control & (1) and (2) & (1) and (3) & (1) and (4) \\
& group & group & group & group & & & \\ \cline{2-5} \cline{6-8}
& (1) & (2) & (3) & (4) &(5) & (6) & (7) \\
\hline\noalign{\smallskip}
\multicolumn{8}{l}{\textbf{Personal characteristics}}\\
\hline\noalign{\smallskip}
Age & 38.8 & 41.3 & 38.7 & 41.5 & 28.5 & 0.9 & 31.4 \\
Older than 50 years & .010 & .111 & .019 & .125 & 43.3 & 7.1 & 47.0 \\
Incapacity (e.g., illness, pregnancy) & .022 & .050 & .032 & .062 & 15.4 & 6.2 & 20.2 \\
Health & .083 & .128 & .093 & .146 & 14.5 & 3.4 & 20.0 \\
\hline\noalign{\smallskip}
\multicolumn{8}{l}{\textbf{Education and occupation}}\\
\hline\noalign{\smallskip}
University entry degree (Abitur) & .229 & .170 & .197 & .142 & 14.7 & 7.9 & 22.5 \\
White-collar & .382 & .476 & .440 & .527 & 19.2 & 12.0 & 29.5 \\
Manufacturing & .069 & .101 & .101 & .147 & 11.7 & 11.4 & 25.3 \\
\hline\noalign{\smallskip}
\multicolumn{8}{l}{\textbf{Employment and welfare history}}\\
\hline\noalign{\smallskip}
Half months empl. (last 2 years) & 45.6 & 44.9 & 44.5 & 43.7 & 10.1 & 15.4 & 25.7 \\
Half months since last unempl. in last 2 years & 46.8 & 46.2 & 45.6 & 44.4 & 11.6 & 19.7 & 35.0 \\
Half months since last OLF (last 2 years) & 45.8 & 44.6 & 44.9 & 43.3 & 15.5 & 12.5 & 29.9 \\
Eligibility unempl. benefits & 13.5 & 14.7 & 13.2 & 14.8 & 21.1 & 5.9 & 20.7 \\
Remaining unempl. insurance claim & 25.6 & 22.3 & 23.4 & 21.4 & 25.0 & 18.0 & 31.7 \\
Cumulative earnings (last 4 years) & 91,204 & 83,632 & 80,913 & 81,156 & 15.6 & 21.8 & 21.0 \\
\hline\noalign{\smallskip}
\multicolumn{8}{l}{\textbf{Timing of unemployment and programme start}}\\
\hline\noalign{\smallskip}
Start unempl. in September & .151 & .079 & .099 & .075 & 22.9 & 15.7 & 24.2 \\
Elapsed unempl. duration & 5.06 & 3.55 & 4.53 & 3.45 & 46.0 & 15.7 & 49.0 \\
\hline\noalign{\smallskip}
\multicolumn{6}{l}{\textbf{Characteristics of local employment agency districts}}\\
\hline\noalign{\smallskip}
Share of empl. in construction industry & .064 & .065 & .077 & .077 & 2.3 & 54.3 & 55.5 \\
Share of male unempl. & .564 & .563 & .541 & .541 & 1.1 & 50.8 & 53.5 \\
\noalign{\smallskip}\hline\hline\noalign{\smallskip}
\end{tabularx}
\parbox{\textwidth}{\scriptsize Note: See Table \ref{descriptives_app} in Online Appendix \ref{app1} for sample first moments of observed characteristics with small standardised differences. In columns (1)-(4), we report the sample first moments of observed characteristics for the treated and non-treated sub-samples. Information on individual characteristics refers to the time of inflow into unemployment, with the exception of the elapsed unemployment duration and monthly regional labour market characteristics, which refer to the (pseudo) treatment time. In columns (5)-(7), we report the standardised differences between the different sub-samples and the treatment group under the voucher system. A description of how we measure absolute standardised differences is available in Online Appendix \ref{app3}. \cite{ro83} classify absolute standardised difference of more than 20 as ``large''. OLF is the acronym for ``out of labour force''.}
\end{center}
\end{table}

%%%%%%%%%%%%%%%%%%%%%
%%%%% TABLE 3 %%%%%%%
%%%%%%%%%%%%%%%%%%%%%

\begin{table}[htb]
\begin{center}
\caption{Average programme durations by training type.}
\renewcommand{\arraystretch}{1} \label{types} \scriptsize
\begin{tabular}{lcccc}
\noalign{\smallskip}\hline\hline\noalign{\smallskip}
& \multirow{2}{*}{\# Obs} & \multirow{2}{*}{Percent} & Average planned & Average actual \\
& & & duration & duration \\ \hline\noalign{\smallskip}
& \multicolumn{4}{c}{Pre-Reform}\\ \cline{2-5}\noalign{\smallskip}
Practice firms  &   11,231 &  16\% &  201 days &  191 days \\
Short training  &   14,564 &  21\% &  114 days &  114 days \\
Long training   &   28,348 &  41\% &  352 days &  336 days \\
Retraining      &   13,340 &  19\% &  762 days &  719 days \\
Others & 1,065  & 2\% & 403 days & 383 days \\ \hline\noalign{\smallskip}
& \multicolumn{4}{c}{Post-Reform}\\ \cline{2-5} \noalign{\smallskip}
Practice firms &    3,409 & 13\% &  156 days & 152 days \\
Short training &   10,864 & 42\% &  116 days & 115 days \\
Long training  &    4,985 & 19\% &  272 days & 279 days \\
Retraining     &    6,487 & 25\% &  799 days & 774 days \\
Others & 590 & 1\% & 467 days & 434 days \\ \noalign{\smallskip}\hline\hline\noalign{\smallskip}
\end{tabular}
\parbox{100mm}{\scriptsize Note: We use the baseline sample (Sample A). The category ''Others'' contains different types of training programmes with very few participants, e.g., programmes that focus on career improvements.}
\end{center}
\end{table}

\clearpage
%\end{document}
%%%%%%%%%%%%%%%%%
%%% APPENDIX %%%%
%%%%%%%%%%%%%%%%%

\renewcommand\appendix{\par
\setcounter{section}{0}%
\setcounter{subsection}{0}%
\setcounter{table}{0}%
\renewcommand\thesection{\Alph{section}}%
\renewcommand\thetable{\Alph{section}.\arabic{table}}}
\renewcommand\thefigure{\Alph{section}.\arabic{figure}}

%%%%%%%%%%%%%%%%%%%%%%%%%%%%%%%%%%%%%%%%%%%%%%%%%%%%%%%%%%%%%%%%%%%%%%%%%%%%%%%%%%%%%%
%%%%%%%%%%%%%%%%%%%%%%%%%%%%%%%%%%%%%%%%%%%%%%%%%%%%%%%%%%%%%%%%%%%%%%%%%%%%%%%%%%%%%%
%%%%%%%%%%%%%%%%%%%%%%%%%%%%%%%%%%%%%%%%%%%%%%%%%%%%%%%%%%%%%%%%%%%%%%%%%%%%%%%%%%%%%%
\clearpage
\setcounter{page}{1}
\setcounter{footnote}{0}
\setcounter{equation}{0}
\onehalfspacing
\begin{center}

{\LARGE Online Appendix to 

``Identifying causal channels of policy reforms with multiple treatments and different types of selection''} \bigskip \bigskip

{\large {Annabelle Doerr and Anthony Strittmatter\footnote{Annabelle Doerr, UC Berkeley, annabelle.doerr@berkeley.edu and Anthony Strittmatter, Department of Economics, University of St.Gallen, anthony.strittmatter@unisg.ch.}}} \bigskip
\end{center}

\subsection*{Sections:}
\begin{enumerate}
\item[A.] Descriptive statistics
\item[B.] Supplements to the empirical approach 
\item[C.] Matching quality 
\item[D.] The change in dropout rates
\item[E.] Results for monthly earnings
\item[F.] Heterogeneous results by programme type
\end{enumerate}

\begin{appendix}
\clearpage
	
%%%%%%%%%%%%%%%%%%%%
%%%% APPENDIX 1 %%%%
%%%%%%%%%%%%%%%%%%%%

\section{Descriptive statistics}\label{app1} \setcounter{table}{0}
\begin{table}[h]
\begin{center}
\centering \caption{Sample first moments of observed characteristics with small standardised differences.}
\renewcommand{\arraystretch}{1}
\label{descriptives_app}\tiny
\begin{tabularx}{160mm}{Xccccccc}\noalign{\smallskip}\hline\hline\noalign{\smallskip}
& \multicolumn{2}{c}{Voucher system} & \multicolumn{2}{c}{Mandatory system}& \multicolumn{3}{c}{Standardised differences between} \\
& Treatment- & Control- & Treatment- & Control- & (1) and (2)  & (1) and (3) & (1) and (4) \\
& group & group & group & group & & & \\ \cline{2-5} \cline{6-8}
& (1) & (2) & (3) & (4) &(5) & (6) & (7) \\ \hline\noalign{\smallskip}
\multicolumn{8}{l}{\textbf{Personal characteristics}}\\ \hline\noalign{\smallskip}
Female                    & .472 &  .447 & .477 & .411 &  5.0 &  .9 & 12.4 \\
No German citizenship     & .054 &  .080 & .052 & .071 & 10.5 & 1.0 &  7.2 \\
Children under 3 years    & .042 &  .035 & .040 & .031 &  3.7 & 1.2 &  6.1 \\
Single                    & .300 &  .285 & .270 & .251 &  3.4 & 6.7 & 11.1 \\
Sanction                  & .007 &  .007 & .009 & .008 &   .2 & 2.0 &   .5 \\
Lack of motivation        & .007 &  .007 & .009 & .008 &   .2 & 2.0 &   .5 \\
\hline\noalign{\smallskip}
\multicolumn{8}{l}{\textbf{Education and occupation}}\\\hline\noalign{\smallskip}
No schooling degree       & .036 &  .068 & .036 & .056 & 14.3 &  .4 & 9.3 \\
Schooling degree without Abitur	 & .720 &  .731 & .750 & .770 & 2.4 & 6.8 & 11.4 \\
Missing                   & .014 & .031 &  .017 & .032 & 10.9 & 2.4 & 11.9 \\
No vocational degree      & .203 & .227 &  .218 & .219 &  5.9 & 3.6 &  3.8 \\
Academic degree           & .112 & .096 &  .081 & .063 &  5.4 &10.6 & 17.3 \\
Agriculture, Fishery      & .012 & .020 &  .015 & .023 &  6.7 & 3.2 &  8.7 \\
Construction              & .054 & .032 &  .027 & .022 & 10.9 &13.3 & 16.6 \\
Trade and Retail          & .127 & .169 &  .148 & .175 & 11.8 & 6.2 & 13.5 \\
Communication and Information Service & .108 & .137 & .122 & .128 & 8.6 & 4.2 & 6.1 \\ \hline\noalign{\smallskip}
\multicolumn{8}{l}{\textbf{Employment and welfare history}}\\
\hline\noalign{\smallskip}
Half months unempl. in last 2 years & .398 &       .370 &       .578 &       .581 &        1.6 &        9.5 &        9.7 \\
No unempl. in last 2 years          & .914 &       .921 &       .877 &       .878 &        2.7 &       11.8 &       11.6 \\
Unemployed in last 2 years          & .034 &       .040 &       .046 &       .052 &        3.1 &        6.2 &        9.1 \\
\# unemployment spells in last 2 years  & .113 &       .102 &       .166 &       .165 &        2.8 &       11.6 &       11.4 \\
Cumulative empl. in last 4 years    & 81.1 &       79.1 &       79.1 &       78.8 &        9.2 &        8.9 &       10.5 \\
Cumulative benefits in last 4 years & 3.00 &       3.52 &       3.70 &       4.02 &        6.3 &        8.2 &       11.5 \\
Any programme in the last 2 years   & .047 &       .042 &       .056 &       .049 &        2.6 &        4.3 &        1.0 \\
\hline\noalign{\smallskip}
\multicolumn{8}{l}{\textbf{Timing of unemployment and programme start}}\\
\hline\noalign{\smallskip}
Start unempl. in January    &  .060 &       .101 &       .117 &       .105 &       15.0 &       19.8 &       16.1 \\
Start unempl. in February   &  .070 &       .089 &       .108 &       .089 &        7.2 &       13.4 &        7.0 \\
Start unempl. in March      &  .096 &       .083 &       .105 &       .085 &        4.5 &        3.0 &        3.7 \\
Start unempl. in April      &  .102 &       .088 &       .120 &       .086 &        4.8 &        5.7 &        5.8 \\
Start unempl. in June       &  .059 &       .078 &       .058 &       .072 &        7.6 &         .6 &        5.3 \\
Start unempl. in July       &  .052 &       .080 &       .053 &       .078 &       11.1 &         .3 &       10.4 \\
Start unempl. in August     &  .081 &       .078 &       .080 &       .078 &        1.0 &         .3 &         .9 \\
Start unempl. in October    &  .127 &       .078 &       .085 &       .082 &       16.4 &       13.8 &       14.9 \\
Start unempl. in November   &  .086 &       .079 &       .045 &       .082 &        2.6 &       16.6 &        1.7 \\
Start unempl. in December   &  .045 &       .082 &       .040 &       .089 &       15.0 &        2.8 &       17.6 \\
\hline\noalign{\smallskip}
\multicolumn{6}{l}{\textbf{State of residence}}\\
\hline\noalign{\smallskip}
Baden-W\"urttemberg                                         & .087 &  .113 & .095 & .090 &  8.6 &  2.9 &  1.2 \\
Bavaria                                                     & .159 &  .138 & .111 & .115 &  6.1 & 14.1 & 12.8 \\
Berlin, Brandenburg                                         & .093 &  .093 & .107 & .111 &  .1  &  4.7 &  6.0 \\
Hamburg, Mecklenburg Western Pomerania, Schleswig Holstein  & .076 &  .088 & .098 & .092 &  4.3 &  7.9 &  5.6 \\
Hesse                                                       & .064 &  .068 & .063 & .058 &  1.7 &   .1 &  2.3 \\
Northrhine-Westphalia                                       & .232 &  .206 & .182 & .197 &  6.2 & 12.4 &  8.6 \\
Rhineland Palatinate, Saarland                              & .056 &  .054 & .055 & .049 &   .9 &   .6 &  3.4 \\
Saxony-Anhalt, Saxony, Thuringia                            & .123 &  .142 & .189 & .190 &  5.5 & 18.4 & 18.5 \\
\hline\noalign{\smallskip}
\multicolumn{6}{l}{\textbf{Characteristics of local employment agency districts}}\\
\hline\noalign{\smallskip}
Population per $km^2$     &       910 &   889 &        789 &        895 &        1.3 &        7.5 &         .9 \\
Unemployment rate (in \%) &      12.2 &  12.3 &       12.1 &       12.0 &        1.9 &        1.4 &        3.8 \\
Share of empl. in production industry &  .250 &       .246 &       .246 &       .241 &        5.1 &        4.7 &        9.9 \\
Share of empl. in trade industry      &  .150 &       .150 &       .150 &       .150 &        1.8 &        2.7 &        2.8 \\
Share of non-German unempl.     &        .139 &       .141 &       .126 &       .128 &        2.5 &       14.3 &       12.1 \\
Share of vacant fulltime jobs   &        .794 &       .794 &       .800 &       .799 &          0 &        8.4 &        7.6 \\
\noalign{\smallskip}\hline\hline\noalign{\smallskip}
\end{tabularx}
\parbox{16cm}{ Note: See Table \ref{descriptives} for sample first moments of observed characteristics with large standardised differences. In columns (1)-(4), we report the sample first moments of observed characteristics for the treated and non-treated sub-samples. Information on individual characteristics refers to the time of inflow to unemployment, with the exception of the elapsed unemployment duration and monthly regional labour market characteristics, which refer to the (pseudo) treatment time. In columns (5)-(7), we report the standardised differences between the different sub-samples and the treatment group under the voucher system. Please find a description of how we measure standardised differences in Online Appendix \ref{app3}. OLF is the acronym for ``out of labour force''.}
\end{center}
\end{table}

\begin{table}[]
\begin{center}
\caption{Efficient first moments of observed characteristics.}
\renewcommand{\arraystretch}{1}\label{balance} \footnotesize\tiny
\begin{tabularx}{150mm}{Xccc}\noalign{\smallskip}\hline\hline\noalign{\smallskip}
& Voucher & Mandatory & Standardised differences between  \\
& system  &  system  & (1) and (2) \\ \cline{2-4}
& (1) & (2) & \\ \hline\noalign{\smallskip}
\multicolumn{4}{l}{\textbf{Personal characteristics}}\\ \hline\noalign{\smallskip}
Female                     & .472 &       .476 &       .8 \\
Age                        & 38.754 &     38.697 &     .8 \\
Older than 50 years        & .011 &       .019 &      7.1 \\
No German citizenship      & .054 &       .052 &        1.1 \\
Children under 3 years     & .042 &       .040 &        1.2 \\
Single                     & .300 &       .270 &        6.6 \\
Health problems            & .083 &       .093 &        3.7 \\
Sanction                   & .007 &       .009 &        2.1 \\
Incapacity (e.g., illness, pregnancy)  & .022 &       .032 &        6.3 \\
Lack of motivation         & .007 &       .009 &        2.1 \\
\hline\noalign{\smallskip}
\multicolumn{4}{l}{\textbf{Education and occupation}}\\
\hline\noalign{\smallskip}				
No schooling degree                       &  .036     & .035  &   .5 \\
Schooling degree without Abitur           &  .719     & .762  &  9.8 \\
University entry degree (Abitur)          &  .230     & .185  & 11.2 \\
No vocational degree                      &  .204     & .217  &  3.3 \\
Academic Degree                           &  .114     & .081  & 11.3 \\
White-collar                              &  .383     & .440  & 11.7 \\
Agriculture, Fishery                      &  .012     & .015  &  3.2 \\
Manufacturing                             &  .069     & .101  & 11.6 \\
Construction                              &  .053     & .027  & 13.1 \\
Trade and Retail                          &  .127     & .148  &  6.1 \\
Communication and Information Service     &  .109     & .122  &  4.1 \\
\hline\noalign{\smallskip}
\multicolumn{4}{l}{\textbf{Employment and welfare history}}\\
\hline\noalign{\smallskip}
Half months empl. in last 2 years               &   45.5  &  44.5  &  15 \\
Half months unempl. in last 2 years             &   .401  &  .587  &  10 \\
Half months since last unempl. in last 2 years  &   46.7  &  45.6  &  20.7 \\
No unempl. in last 2 years                      &   .913  &  .876  &  12.1 \\
Unempl. in last 2 years                         &   .034  &  .047  &  6.3 \\
\# unemployment spells in last 2 years          &   .114  &  .168  &  12 \\
Any programme in last 2 years                     &   .047  &  .057  &  4.5 \\
Half months since of last OLF in last 2 years   &   45.7  &  44.9  &  12.1 \\
Remaining unempl. insurance claim               &   25.6  &  23.4  &  17.6 \\
Eligibility unempl. benefits                    &   13.5  &  13.2  &  5.9 \\
Cumulative empl. in last 4 years                &   81    &  79    &  8.3 \\
Cumulative earnings in last 4 years             &   91,018&  80,928&  21.1 \\
Cumulative benefits in last 4 years             &   3.02  &  3.74  &  8.4 \\	
\hline\noalign{\smallskip}
\multicolumn{4}{l}{\textbf{Timing of unemployment and programme start}}\\
\hline\noalign{\smallskip}
Start unempl. in January    &       .059 & .116  &       19.9 \\
Start unempl. in February   &       .070 & .108  &       13.4 \\
Start unempl. in March      &       .095 & .104  &        2.9 \\
Start unempl. in April      &       .102 & .120  &        5.7 \\
Start unempl. in June       &       .059 & .058  &         .4 \\
Start unempl. in July       &       .052 & .053  &         .7 \\
Start unempl. in August     &       .082 & 0.08  &         .6 \\
Start unempl. in September  &       .152 & .099  &       16.2 \\
Start unempl. in October    &       .127 & .085  &       13.5 \\
Start unempl. in November   &       .087 & .046  &       16.6 \\
Start unempl. in December   &       .045 & .040  &        2.6 \\
Elapsed unempl. duration    &       5.08 & 4.54  &       16.2 \\
\hline\noalign{\smallskip}
\multicolumn{4}{l}{\textbf{State of residence}}\\
\hline\noalign{\smallskip}
Baden-W\"urttemberg    &       .085  &       .093 &        2.9 \\
Bavaria                &       .159  &       .113 &       13.4 \\
Berlin, Brandenburg    &       .090  &       .103 &        4.3 \\
Hamburg, Mecklenburg Western Pomerania, Schleswig Holstein & .077 & .099 & 8\\                                 
Hesse                  &       .064  &       .064 &          0 \\
Northrhine-Westphalia                &       .231  &       .180 &       12.7 \\
Rhineland Palatinate, Saarland       &       .056  &       .055 &         .7 \\
Saxony-Anhalt, Saxony, Thuringia     &       .125  &       .191 &         18 \\
\hline\noalign{\smallskip}
\multicolumn{4}{l}{\textbf{Characteristics of local employment agency districts}}\\
\hline\noalign{\smallskip}
Share of empl. in production industry     &   .250 &  .246     &     5.1 \\
Share of empl. in construction industry   &   .064 &  .077     &    52.4 \\
Share of empl. in trade industry          &   .150 &  .150     &     3.1 \\
Share of male unempl.                     &   .564 &  .541     &    46.9 \\
Share of non-German unempl.               &   .138 &  .126     &    13.3 \\
Share of vacant fulltime jobs             &   .793 &  .800     &     8.9 \\
Population per $km^2$                     &    902 &   778     &     7.4 \\
Unemployment rate (in \%)                 &   12.2 &  12.1     &     2.5 \\\noalign{\smallskip}\hline\hline\noalign{\smallskip}
\end{tabularx}
\parbox{16cm}{ Note: In columns (1)-(2), we report the efficient first moments of observed characteristics for the treated sub-samples. They are exactly equal in the other re-weighted sub-samples, which are not reported. Information on individual characteristics refers to the time of inflow to unemployment, with the exception of the elapsed unemployment duration and monthly regional labour market characteristics which refer to the (pseudo) treatment time. In column (3), we report the standardised differences (SD) between the two treatment groups. Please find a description of how we measure standardised differences in Online Appendix \ref{app3}. OLF is the acronym for ``out of labour force''.}
\end{center}
\end{table}
	
\clearpage
	
%%%%%%%%%%%%%%%%%%%%
%%%% APPENDIX B %%%%
%%%%%%%%%%%%%%%%%%%%

\section{Supplements to the empirical approach}

\subsection{Proof of Equation (\ref{proof1}) \label{app2}}
We show that $E[Y^d_{i,t}(s)|D_i=g, T_i=q]$ can be identified from the joint
distribution of random variables $(Y,G(d,t,s),G(d',t',s),X)$ under Assumptions
1a and 2a \citep[comp.][]{hi03,ro83}:
\begin{align*}
E[Y^d_{i,t}(s)|D_i=d', T_i=t'] = &\int E[Y^d_{i,t}(s)|D_i=d',T_i=t',X_i=x] f_X(x|D_i=d', T_i=t') dx,\\
=&\int E[Y^d_{i,t}(s)|D_i=d,T_i=t,X_i=x] f_X(x|D_i=d', T_i=t') dx,\\
=&\int E[Y_i|D_i=d,T_i=t,X_i=x] f_X(x|D_i=d', T_i=t')dx,\\
=&\int E[G_i(d,t,s)Y_i|D_i=d,T_i=t,X_i=x] f_X(x|D_i=d', T_i=t')dx,\\
=&\int \frac{1}{p_{d,t,s}(x)} E[G_i(d,t,s)Y_i|X_i=x] f_X(x|D_i=d', T_i=t') dx,\\
=&\int \frac{p_{d',t',s}(x)}{p_{d',t',s}\cdot p_{d,t,s}(x)} E[G_i(d,t,s)Y_i|X_i=x]f_X(x)dx ,\\
=&\int \frac{p_{d',t',s}(x)}{p_{d',t',s}\cdot p_{d,t,s}(x)} G_i(d,t,s)Y_i f_X(x) dx ,\\
=&E\left[\frac{p_{d',t',s}(x)}{p_{d',t',s}\cdot p_{d,t,s}(x)} G_i(d,t,s)Y_i \right] .
\end{align*}
In the first equation we apply the law of iterative expectations. In the second equality we condition on $D_i=d$, which is possible
because we assume that the expected potential outcomes are independent of the treatment after controlling for $X_i$ (Assumption 1). In equality three we replace the potential by the observed outcome. In equality four we multiply the outcome $Y_i$ with the the group dummy $G_i(d,t,s)$. In equality five we use the fact that $E[DY]=E[DY|D=1]Pr(D=1)$. In equality six we apply Bayes' rule. We make a backward application of the law of iterative expectations in equality seven. Finally, we replace the integral by an expectation in equality eight.
\begin{flushright}
\vspace*{-1cm}$\Box$
\end{flushright}

\clearpage 

%%%%%%%%%%%%%%%%%%%%
%%%% APPENDIX C %%%%
%%%%%%%%%%%%%%%%%%%%

\subsection{Estimation strategy \label{app3}}

A straightforward estimation strategy is based on the sample analogue
of (\ref{proof1})
\begin{equation*} \label{w2}
\hat{E}[Y^d_{i,t}(s)|D_i=d', T_i=t'] = \frac{1}{N} \sum_{i=1}^{N} \hat{\omega}_iY_i,
\end{equation*}
with
\begin{equation} \label{w1}
\hat{\omega}_i= \frac{G_i(d,t,s)}{\frac{1}{N}\sum_{j=1}^{N} \hat{p}_{d',t',s}(X_j)} \cdot \frac{\hat{p}_{d',t',s}(X_i)}{\hat{p}_{d,t,s}(X_i)},
\end{equation}
where $\hat{p}_{d',t',s}(X_i)$ and $\hat{p}_{d,t,s}(X_i)$ indicate the estimated conditional treatment probabilities (henceforth, \emph{propensity scores}). This is an \emph{Inverse Probability Weighting} (IPW) estimator. \cite{hi03} demonstrate that the consistency and efficiency of an IPW critically depend on the estimated propensity scores. Parametric specifications of the propensity score do not necessarily lead to efficient estimates. One reason is that (\ref{w1}) seeks to balance the sample covariate distributions, which equal
\begin{equation*}
\hat{F}_{d',t'} = \frac{1}{\sum_{i=1}^{N}\hat{p}_{d',t',s}(X_i)}\sum_{i=1}^{N}
G_i(d',t',s)1\{X_i \leq x\},
\end{equation*}
when $d=d'$ and $t=t'$. However, $\hat{F}_{d',t'}$ can be more efficiently estimated using information from the entire population rather than from the random sample $d',t'$ alone. The efficient estimators for the covariate distributions of subpopulation $d',t'$ equal
\begin{equation*}
\hat{F}_{d',t'}^{eff}= \frac{1}{\sum_{i=1}^{N}
	\hat{p}_{d',t',s}(X_i)}\sum_{i=1}^{N}\hat{p}_{d',t',s}(X_i)1\{X_i \leq x\}.
\end{equation*}
Accordingly, reweighting estimators that recover $\hat{F}_{d',t'}^{eff}$ rather than of $\hat{F}_{d',t'}$ may be more efficient. We report the efficient first moments for all control variables and both treatment groups in Table \ref{balance} in Online Appendix \ref{app1}.

\cite{gr16} recently proposed a double robust and locally efficient semiparametric version of IPW, named \emph{Auxiliary-to-Study Tilting} (AST). This estimator precisely balances the efficient first moments of all control variables in each treatment sample.\footnote{Exact balancing is not guaranteed for the sample moments using conventional IPW estimators.} Using AST, the propensity score is estimated in a conventional parametric way. We use the probit model $\hat{p}_{d',t',s}(X_i)=\Phi(X_i' \hat{\beta})$, where $\Phi(\cdot)$ denotes the cumulative normal distribution function and $X_i' \hat{\beta}$ is the estimated linear index. The estimated propensity score $\hat{p}_{d,t,s}(x)$ is replaced by $\tilde{p}_{d,t,s}(x)$. It is estimated under the following moment conditions:
\begin{equation}\label{ast1}
\frac{1}{N} \sum_{i=1}^{N}
\frac{\displaystyle G_i(d,t,s)}{\displaystyle \frac{1}{N} \sum_{j=1}^{N} \hat{p}_{d',t',s}(X_j)}  \cdot \frac{\displaystyle\hat{p}_{d',t',s}(X_i)}{\displaystyle \tilde{p}_{d,t,s}(X_i)} \cdot X_i
=
\displaystyle  \frac{1}{N} \sum_{i=1}^{N}  \frac{\displaystyle\hat{p}_{d',t',s}(X_i)}{\displaystyle \frac{1}{N} \sum_{j=1}^{N} \hat{p}_{d',t',s}(X_j)} \cdot X_i,
\end{equation}
where $\tilde{p}_{d,t,s}(X_i)=\Phi ( X_i' \tilde{\beta} )$ is specified such that the left and right sides of (\ref{ast1}) are numerically equivalent for all elements in $X_i$ (including a constant term). The right side is the efficient first moment estimate. As the efficient first moment estimates are independent of subpopulation with $d,t$, the first moments are exactly balanced in all treatment groups for $d, t \in \{0,1\}$ using this procedure. The constant guarantees that the weights sum to one. The expected potential outcomes are estimated using
\begin{equation*}
\tilde{E}[Y^d_{i,t}(s)|D_i=d', T_i = t'] = \frac{1}{N} \sum_{i=1}^{N}
\tilde{\omega}_iY_i ,
\end{equation*}
with
\begin{equation*}
\tilde{\omega}_i= \frac{G_i(d,t,s)}{\frac{1}{N}\sum_{j=1}^{N} \hat{p}_{d',t',s}(X_j)} \cdot \frac{\hat{p}_{d',t',s}(X_i)}{ \tilde{p}_{d,t,s}(X_i)}.
\end{equation*}
It can be shown that this estimator is $\sqrt{N}$-consistent and asymptotically normal distributed.\footnote{The large sample properties of AST are subject to assumptions regarding the specification of the propensity score. These assumptions imply that the propensity score is correctly specified, strictly increasing in its arguments, differentiable, and well located within the unit interval.} Similar to \cite{gr16}, we compute the significance levels (p-values) of our estimated parameters based on a non-parametric bootstrapping procedure (sampling individual observations with replacement).

\clearpage

%%%%%%%%%%%%%%%%%%%%
%%%% APPENDIX D %%%%
%%%%%%%%%%%%%%%%%%%%

\section{Matching quality \label{app4}} \setcounter{table}{0}
We assess the matching quality by reporting the moments (mean, variance, skewness, kurtosis) and standardised differences for the control variables in all four samples.  The standardised differences
are defined by
\begin{equation*}
SD = \frac{\left| \mu_{d,t,s} - \mu_{d',t',s} \right|}{\sqrt{0.5 (\sigma_{\mu_{d,t,s}}^2 + \sigma_{\mu_{d',t',s}}^2)}} \cdot 100\%,
\end{equation*}
where $\mu_{d,t,s}$ is the moment and $ \sigma_{\mu_{d,t,s}}^2$ is the variance of the moment in the respective treatment group $G_i(d,t,s)$ with $d, d', t, t' \in\{0,1\}$ and $s \in\{v,m\}$. The pre-matching standardised differences between the sample first moments are reported in Table \ref{descriptives}. The post-matching standardised differences between the efficient first moments are exactly zero, as the first moments are precisely balanced (see the discussion in Online Appendix \ref{app3}). Therefore, we do not report the standardised difference of the matched treatment and control samples in Table \ref{balance} (only between the voucher and mandatory system).

In the optimal case, matching estimators balance the complete distributions of all control variables rather than only the first moments. For all binary variables, this requirement is satisfied because the first moments are balanced. In the main specifications, we control for 63 variables, 43 of which are binary. For the other variables, we report the variance, skewness, and kurtosis for the different samples matched to the treatment group under the voucher system in Table \ref{moment1}. Furthermore, we present the higher moments for the different samples matched to the treatment group under the mandatory system in Table \ref{moment2}. For most moments, we report small standardised differences. However, particularly for the monthly regional labour market characteristics, we find large differences in the higher moments for the samples that are matched to the treatment group under the mandatory system.

\clearpage

\begin{table}[]
\begin{center}
\centering \caption{Higher moments of observed characteristics matched to the treatment group under the voucher system.}
\renewcommand{\arraystretch}{1}
\scriptsize \label{moment1}
\resizebox{160mm}{!}{\begin{tabular}{lccccccc}
\noalign{\smallskip}\hline\hline\noalign{\smallskip}
& \multicolumn{2}{c}{Voucher system} &
\multicolumn{2}{c}{Mandatory system}& \multicolumn{3}{c}{Standardised differences between}  \\
& Treatment-  & Control- & Treatment- & Control- & (1) and (2) & (1) and (3) & (1) and (4)  \\
& group & group & group & group & & & \\ \cline{2-5} \cline{6-8}
& (1) & (2) & (3) & (4) &(5) & (6) & (7) \\
\hline\noalign{\smallskip}
\multicolumn{8}{c}{\textbf{Variance}} \\
\hline\noalign{\smallskip}
Age & 55.48 & 64.64 & 56.33 & 63.1 & 13.62 & 1.35 & 11.68 \\
Half months empl. in the last 24 months & 41.45 & 40.17 & 38.23 & 36.72 & 1.16 & 2.98 & 4.47 \\
Half months unempl.in the last 24 months & 2.98 & 3.12 & 3.12 & 3.14 & .72 & .67 & .77 \\
Time since last unemployment in the last 24 months (half-months) & 20.07 & 20.45 & 20.73 & 20.5 & .42 & .67 & .44 \\
\# unemployment spells in the last 24 months & .17 & .17 & .17 & .17 & .20 & .07 & .48 \\
Time of last out of labour force in last 24 months & 42.36 & 41.97 & 40.48 & 40.62 & .26 & 1.34 & 1.24\\
Remaining unemployment insurance claim & 174.91 & 193.14 & 163.49 & 185.54 & 6.64 & 4.47 & 4.00 \\
Eligibility unemployment benefits & 25.6 & 25.53 & 25.83 & 25.22 & .17 & .54 & .89 \\
Cumulative employment (last 4 years before unemployment) &513.34 & 468.74 & 491.27 & 442.71 & 5.73 & 2.83 & 9.28 \\
Cumulative earnings (last 4 years before unemployment)   & 2.42 $ \cdot 10^9 $&2.52 $ \cdot 10^9$&2.38 $ \cdot 10^9 $ & 2.48 $\cdot 10^9$ & 2.71 & 1.33 & 1.68 \\
Cumulative benefits (last 4 years before unemployment) & 62.67 & 61.98 & 66.42 & 66.77 & .23 & 1.19 & 1.22 \\
Elapsed unemployment duration                      & 11.31 & 12.23 & 13.06 & 12.4  & 8.58&16.09 & 10.22 \\
Share of empl.in production industry               &.00783 &.00816 &.00913 &.00918 &3.54 &12.84 & 13.40 \\
Share of empl. in construction industry            &.0004  &.00043 &.00038 &.00039 &4.26 & 3.82 &  1.98 \\
Share of empl. in trade industry 				   &.00032 &.00035 &.00032 &.00030 &6.03 & 1.03 &  3.65 \\
Share of male unempl.                              &.00175 &.00175 &.00154 &.00152 & .15 & 8.55 &  9.64 \\
Share of non-German unempl.                        &.00734 &.00742 &.00659 &.00644 & .97 & 9.16 & 10.92 \\
Share of vacant fulltime jobs                      &.00584 &.00583 &.00383 &.00391 & .12 &21.42 & 20.81 \\
Population per $km^2$                              &2814047&2820432 & 2039321 &2152881 &.07 &8.78 &7.37 \\
Unemployment rate (in \%)                          & 25.56 & 26.3 & 19.42 & 19.02 & 2.15 & 19.77 & 20.87 \\
\hline\noalign{\smallskip}
\multicolumn{8}{c}{\textbf{Skewness}}\\
\hline\noalign{\smallskip}
Age & 46.23 & 110.42 & 81.03 & 115.84 & 5.20 & 3.18 & 5.87 \\
Half months empl. in the last 24 months & -691.31 & -663.08 & -610.47 & -566.93 & 1.18 & 3.50 & 5.52 \\
Half months unempl. in the last 24 months & 29.37 & 32.61 & 34.63 & 35.01 & .97 & 1.40 & 1.52 \\
Time since last unemployment in the last 24 months (half-months) & -381.07 & -400.22 & -449.84 & -439.91 & .89 & 2.61 & 2.30 \\
\# unemployment spells in the last 24 months & 0.31 & 0.29 & 0.31 & 0.33 & .59 & .16 & .63 \\
Time of last out of labour force in last 24 months & -865.15 & -866.25 & -789.08 & -797.59 & .03 & 1.80 & 1.61 \\
Remaining unemployment insurance claim  & 1.07 & 353.20 & -210.13 & 48.83 & 3.62 &  2.44 & .52 \\
Eligibility unemployment benefits & 145.11 & 157.14 & 154.61 & 152.87 & 2.10 & 1.72 & 1.33 \\
Cumulative employment (last 4 years before unemployment) & -1.57$ \cdot 10^4 $ & -1.37$ \cdot 10^4 $ & -1.42$ \cdot 10^4 $ & -1.25$ \cdot 10^4 $ & 4.64 & 3.41 & 7.38 \\
Cumulative earnings (last 4 years before unemployment) & 7.04$\cdot 10^{13}$ & 9.0$\cdot 10^{13}$ & 7.02$\cdot 10^{13}$ & 9.41$\cdot 10^{13}$ & 4.15 & 0.04 & 5.03 \\
Cumulative benefits (last 4 years before unemployment) & 2,014.09 & 2,016.49 & 2,268.74 & 2,441.50 & .01 & 1.34 & 1.86\\
Elapsed unemployment duration & 4.15 & 3.54 & 4.58 & 3.41 & .75 & .52 & .91 \\
Share of empl. in production industry   & .00022 & .00031 & .00053 & .00054 & 4.33 &13.62 & 13.83 \\
Share of empl. in construction industry & .00001 & .00001 & .00001 & .00001 & 2.61 & 9.02 & 10.29 \\
Share of empl. in trade industry        & .00000 & .00000 & .00000 & .00000 & 1.20 & 9.53 & 10.75 \\
Share of male unempl.                   &-.00004 &-.00004 & .00000 &-.00001 & .32  &13.85 & 10.23 \\
Share of non-German unempl.             & .00016 & .00016 & .00010 & .00012 & .36  & 3.13 &  2.28 \\
Share of vacant fulltime jobs           &-.00040 &-.00043 &-.00019 &-.00017 &  .92 & 8.39 &  9.28 \\
Population per $km^2$ & 1.52$ \cdot 10^{10} $ & 1.53$ \cdot 10^{10} $ & 1.03$ \cdot 10^{10} $ & 1.11$ \cdot 10^{10} $ & 0.17 & 8.30 & 6.82 \\
Unemployment rate (in \%) & 112.15 & 124.58 & 71.48 & 69.47 & 2.82 & 10.81 & 11.08 \\
\hline\noalign{\smallskip}
\multicolumn{8}{c}{\textbf{Kurtosis}} \\
\hline\noalign{\smallskip}
Age & 6984 & 9302 & 7132 & 8593 & 13.28 & 1.05 & 9.75 \\
Half months empl. in the last 24 months & 14214 &  13745 & 12377 & 11302 & .88 & 3.64 & 5.95\\
Half months unempl. in the last 24 months & 375 & 440 & 521 & 520 & .96 & 1.82 & 1.84\\
Time since last unemployment in the last 24 months (half-months) & 8409 & 9053 & 11762 & 11308 & 1.22 & 4.30 & 3.95\\
\# unemployment spells in the last 24 months  & .84 & .69 & .75 & .89 & .81 & .47 & .27\\
Time of last out of labour force in last 24 months & 22815 & 23205 & 20015 & 20243 & .26 &  1.98 & 1.85 \\
Remaining unemployment insurance claim  & 100847 & 117898 & 87180 & 105122 & 5.70 & 5.40 & 1.60 \\
Eligibility unemployment benefits & 2374 & 2615 & 2533 & 2621 & 3.11 & 2.23 & 3.06 \\
Cumulative employment (last 4 years before unemployment) & 912559 & 781159 & 807997 & 705710 & 5.50 & 4.38 & 9.00 \\
Cumulative earnings (last 4 years before unemployment) & 1.78$ \cdot 10^{19} $ & 2.04$ \cdot 10^{19} $ & 1.74$ \cdot 10^{19} $ & 1.96$ \cdot 10^{19} $ & 4.15 & 0.88 & 3.08 \\
Cumulative benefits (last 4 years before Unemployment) & 94493 & 98694 & 113972 & 139252 & .33 & 1.48 & 2.36 \\
Elapsed unemployment duration & 241 & 266 & 295 & 269 & 7.26 & 15.21 & 8.26 \\
Share of empl. in production industry & .0001413 & .0001612 & .0002073 & .0002076 & 4.96 & 14.45 & 14.70 \\
Share of empl. in construction industry & .0000005 & .0000005 & .0000006 & .0000006 & 4.02 & 5.82 & 8.46 \\
Share of empl. in trade industry & .0000003 & .0000004 & .0000003 & .0000003 & 4.92 & 3.75 & 4.30 \\
Share of male unempl.& .0000095 & .0000098 & .0000078 & .0000074 & 1.03 & 5.88 & 7.43 \\
Share of non-German unempl. & .0001247 &  .0001258 &.000105 & .0001041 & .38 & 7.28 & 7.60 \\
Share of vacant fulltime jobs & .0001584 & .0001677 & .0000669 & .0000632 & .94 & 12.29 & 13.13 \\
Population per $km^2$ & 1.0$\cdot 10^{14}$ & 1.0$ \cdot 10^{14}$ & 6.8$\cdot 10^{14}$ & 7.3$\cdot 10^{14}$ & .22 & 8.46 & 6.91 \\
Unemployment rate (in \%) & 1740 & 1941 & 1219 & 1239 & 3.82 & 12.50 & 11.49 \\ 
\noalign{\smallskip}\hline\hline\noalign{\smallskip}
\end{tabular}}
\parbox{16cm}{Note: In columns (1)-(4), we report the variance, skewness, and kurtosis of observed characteristics for the treated and non-treated sub-samples. Information on individual characteristics refers to the time of inflow into unemployment, with the exception of the elapsed unemployment duration and monthly regional labour market characteristics, which refer to the (pseudo) treatment time. In columns (5)-(7), we report the standardised differences between the different sub-samples and the treatment group under the voucher system. All control variables that are not reported in this table have binary distributions. The higher moments of these variables are precisely balanced in the matched samples.}
\end{center}
\end{table}
	
\clearpage
\begin{table}[]
\begin{center}
\centering \caption{Higher moments of observed characteristics matched to the treatment group under the mandatory system.}
\renewcommand{\arraystretch}{1}\scriptsize \label{moment2}
\resizebox{160mm}{!}{\begin{tabular}{lccccccc}
\noalign{\smallskip}\hline\hline\noalign{\smallskip}
& \multicolumn{2}{c}{Voucher system} & \multicolumn{2}{c}{Mandatory system}& \multicolumn{3}{c}{Standardised differences between}  \\
& Treatment- & Control- & Treatment- & Control- & (1) and (2)  & (1) and (3) & (1) and (4) \\
& group  & group & group & group  & & &  \\ \cline{2-5} \cline{6-8}
& (1) & (2) & (3) & (4) &(5) & (6) & (7) \\
\hline\noalign{\smallskip}
\multicolumn{8}{c}{\textbf{Variance}} \\
\hline\noalign{\smallskip}
Age & 59.75 & 60.3 & 72.45 & 66.49 & .82 & 16.27 & 8.64 \\
Half months empl. in the last 24 months & 58.91 & 54.18 & 62.6 & 53.03 & 3.98 & 6.51 & 1 \\
Half months unempl. in the last 24 months & 3.93 & 4.32 &  4.4 & 4.2 & 1.83 & .33 & .51 \\
Time since last unemployment in the last 24 months (half-months)& 41.11 & 45.83 & 44.56 & 44.64 & 3.34 & .84 & .75 \\
\# unemployment spells in the last 24 months &      .25 &       .26 &       .25 &       .26 &       .64 &       .55 &       .13 \\
Time of last out of labour force in last 24 months &      60.4 &      57.71 &      60.05 &      58.79 &       1.59 &       1.46 &       .66 \\
Remaining unemployment insurance claim                          &    156.52 &     145.13 &     178.65 &     169.72 &       4.91 &      14.17 &       10.4 \\
					Eligibility unemployment benefits                               &     27.96 &      28.14 &      28.14 &      28.02 &       .39 &       .02 &       .24 \\
					Cumulative employment (last 4 years before unemployment)        &    606.44 &     559.64 &     553.45 &     518.33 &       5.73 &       .78 &       5.39 \\
					Cumulative earnings (last 4 years before unemployment)          &  2.14$\cdot 10^{9}$ &   2.04$\cdot 10^{9}$ &   2.11$\cdot 10^{9}$ &   2.15$\cdot 10^{9}$  & 3.152 & 2.242 & 3.228 \\
					Cumulative benefits (last 4 years before unemployment)          &     85.48 &      84.85 &      76.86 &      83.58 &       .17 &       2.28 &       .33 \\
					Elapsed unemployment duration                                   &     10.82 &      12.12 &      13.18 &      12.03 &       11.8 &       9.35 &       .81 \\
					Share of empl. in production industry                     &   .00577 &     .0086 &    .00573 &    .00842 &      32.26 &      31.97 &       1.76 \\
					Share of empl. in construction industry                  &   .00082 &     .0007 &    .00083 &    .00075 &      15.05 &      16.32 &       6.07 \\
					Share of empl. in trade industry                         &   .00041 &    .00038 &    .00038 &    .00038 &       4.26 &       1.29 &       .29 \\
					Share of male unempl.                                       &   .00283 &    .00204 &    .00264 &    .00211 &       26.2 &       21.6 &       .43 \\
					Share of non-German unempl.                                  &   .00991 &     .0084 &    .00931 &    .00833 &      14.99 &       9.19 &       .83 \\
					Share of vacant fulltime jobs                                   &   .00502 &    .00524 &      .005 &    .00485 &        2.3 &       2.57 &       4.11 \\
					Population per $km^2$                                           &   2800928 &    2321975 &    2881181 &    2377444 &       5.02 &       5.81 &       .62 \\
					Unemployment rate (in \%)                                       &     31.81 &      32.41 &      34.03 &      31.58 &       1.99 &       5.25 &       2.52 \\

					\hline\noalign{\smallskip}
					\multicolumn{8}{c}{\textbf{Skewness}} \\
					\hline\noalign{\smallskip}
					
					Age                                                                       &            85.74 &     110.07 &     153.47 &     163.78 &          2 &       2.96 &        3.9 \\
					Half months empl. in the last 24 months                                &          -894.18 &    -795.24 &   -1055.27 &    -772.03 &       3.87 &       8.51 &        .9 \\
					Half months unempl. in the last 24 months                              &            31.99 &      43.57 &      39.91 &      40.72 &       3.36 &       1.02 &       .69 \\
					Time since last unemployment in the last 24 months (half-months)          &           -732.4 &   -1014.77 &    -896.02 &    -969.59 &       7.23 &       2.76 &       .96 \\
					\# unemployment spells in the last 24 months                              &             .44 &       .48 &       .43 &        .5 &       1.09 &       1.44 &       .41 \\
					Time of last out of labour force in last 24 months                        &         -1244.94 &   -1115.88 &   -1147.04 &   -1168.59 &       2.46 &       .66 &       1.05 \\
					Remaining unemployment insurance claim                                    &            126.4 &    -109.32 &      176.5 &     228.35 &       3.05 &       3.59 &        4.3 \\
					Eligibility unemployment benefits                                         &           152.86 &     166.34 &     184.87 &     175.92 &       2.08 &       2.49 &       1.31 \\
					Cumulative employment (last 4 years before unemployment)                  &        -18075.06 &  -15691.71 &  -15693.52 &  -14064.85 &       5.02 &      .004 &       3.72 \\
					Cumulative earnings (last 4 years before unemployment)                    &         6.47$\cdot 10^{13}$ &   6.33$\cdot 10^{13}$ &   7.69$\cdot 10^{13}$ &   8.62$\cdot 10^{13}$ & .34 & 3.08 &       5.03 \\
					Cumulative benefits (last 4 years before unemployment)                    &          3005.55 &     2948.6 &    2372.48 &    2995.96 &       .25 &       2.75 &       .18 \\
					Elapsed unemployment duration                                             &             9.71 &      12.41 &      11.45 &      11.55 &       3.23 &       1.07 &          1 \\
					Share of empl. in production industry                              &        .0001537 &   .000418 &  .0001786 &  .0004147 &      13.62 &      11.99 &       .14 \\
					Share of empl. in construction industry                            &        .0000071 &  .0000112 &  .0000112 &  .0000135 &        8.1 &       .06 &       4.02 \\
					Share of empl. in trade industry                                   &        .0000054 &  .0000044 &  .0000049 &  .0000045 &       3.59 &        1.7 &       .44 \\
					Share of male unempl.                                               &       -.0000996 & -.0000117 & -.0000718 & -.0000223 &      23.68 &      18.35 &        3.3 \\
					Share of non-German unempl.                                           &        .0005108 &  .0002599 &  .0004424 &  .0002689 &         11 &       8.29 &       .44 \\
					Share of vacant fulltime jobs                                             &       -.0002477 & -.0004009 & -.0002327 & -.0003027 &       5.25 &       6.01 &       3.66 \\
					Population per $km^2$                                                     &         1.63$\cdot 10^{10}$ &   1.28$\cdot 10^{10}$ &   1.68$\cdot 10^{10}$ &   1.33$\cdot 10^{10}$ &5.56 & 6.26 &  .84 \\
					Unemployment rate (in \%)                                                 &            104.8 &     104.26 &     115.28 &     101.21 &        .14 &        2.7 &        0.7 \\
					
					\hline\noalign{\smallskip}
					\multicolumn{8}{c}{\textbf{Kurtosis}} \\
					\hline\noalign{\smallskip}
					
					Age                                                                        &           7962.29 &    8231.21 &      11811 &   10104.53 &       1.62 &      15.71 &       8.85 \\
					Half months empl. in the last 24 months                                 &          18211.33 &   16415.04 &   23904.57 &   15972.42 &       3.16 &        9.7 &        .7 \\
					Half months unempl. in the last 24 months                               &             331.3 &     602.69 &     439.63 &     540.75 &       3.88 &       2.29 &       .72 \\
					Time since last unemployment in the last 24 months (half-months)           &          15975.52 &   27816.77 &   22022.23 &   26122.82 &      10.04 &        4.4 &       1.13 \\
					\# unemployment spells in the last 24 months                                &              1.11 &       1.28 &          1 &        1.4 &        .86 &       2.17 &       .69 \\
					Time of last out of labour force in last 24 months                         &          34620.56 &   29581.17 &   28946.72 &   31456.16 &       2.86 &        .41 &       1.12 \\
					Remaining unemployment insurance claim                                     &          83697.48 &   69367.07 &   95652.58 &   92332.71 &       6.03 &      11.01 &       9.91 \\
					Eligibility unemployment benefits                                          &           2813.09 &    3052.54 &    3399.18 &     3331.8 &       2.61 &       2.98 &       2.49 \\
					Cumulative employment (last 4 years before unemployment)                   &           1086431 &   929914.3 &   939481.9 &   827675.9 &       6.41 &        .41 &       4.58 \\
					Cumulative earnings (last 4 years before unemployment)                     &          1.48$\cdot 10^{19}$ &   1.40$\cdot 10^{19}$ &   1.57$\cdot 10^{19}$ &   1.72$\cdot 10^{19}$ &       1.57 &        2.9 &       5.35 \\
					Cumulative benefits (last 4 years before unemployment)                     &          149761.7 &   151238.6 &   108257.2 &   165057.4 &         .1 &       2.98 &        .66 \\
					Elapsed unemployment duration                                              &            232.13 &     274.49 &     302.67 &     268.76 &      10.38 &       6.53 &       1.36 \\
					Share of empl. in production industry                                &           .00008 &    .00018 &    .00009 &    .00018 &       24.78 &      22.17 &        .09 \\
					Share of empl. in construction industry                             &          .0000012 &   .0000011 &   .0000014 & .0000014 &       4.01 &       9.89 &        6.4 \\
					Share of empl. in trade industry                                    &          .0000004 &   .0000004 &   .0000004 &   .0000005 &       1.97 &       1.76 &       2.03 \\
					Share of male unempl.                                                   &           .00002 &    .00001 &    .00002 &    .00001 &      23.47 &      15.54 &       4.27 \\
					Share of non-German unempl.                                          &           .00022 &    .00015 &     .0002 &    .00015 &      16.55 &      11.75 &       .31 \\
					Share of vacant fulltime jobs                                              &           .00012 &    .00013 &    .00011 &     .0001 &        1.8 &       2.87 &        3.6 \\
					Population per $km^2$                                                      &          1.10$\cdot 10^{14}$ &   8.52$\cdot 10^{13}$ &   1.14$\cdot 10^{14}$ &   8.88$\cdot 10^{13}$ &       5.78 &       6.51 &       .89 \\
					Unemployment rate (in \%)                                                  &           1761.13 &    2128.73 &    1985.96 &    2093.44 &       9.62 &       3.59 &        .8 \\
\noalign{\smallskip}\hline\hline\noalign{\smallskip}
\end{tabular}}
\parbox{16cm}{Note: In columns (1)-(4), we report the variance, skewness, and kurtosis of observed characteristics for the treated and non-treated sub-samples. Information on individual characteristics refers to the time of inflow into unemployment, with the exception of the elapsed unemployment duration and monthly regional labour market characteristics, which refer to the (pseudo) treatment time. In columns (5)-(7), we report the standardised differences between the different sub-samples and the treatment group under the voucher system. All control variables that are not reported in this table have binary distributions. The higher moments of these variables are precisely balanced in the matched samples.}
\end{center}
\end{table}
		
\clearpage

%%%%%%%%%%%%%%%%%%%%%
%%%% Appendix E %%%%%
%%%%%%%%%%%%%%%%%%%%%
\section{The change in dropout rates \label{app_dropout}} \setcounter{table}{0} \setcounter{figure}{0}

In our interpretation of the negative effects of voluntary participation over the short- and medium-term after course start (comp. Section \ref{sec362:channels}), we argue that participants might change their attitudes towards training in a positive way and participate with higher motivation. If an increase in motivation actually occurs, we should see a lower dropout rate under the voucher system compared to the mandatory system. Therefore, we implement a simple descriptive analysis to investigate the change in dropout rates under both allocation systems. Course completion or dropout is only observed for treated individuals. We define dropout as proposed by \cite{paul15} if particiants complete less than 80\% of the planned course duration.\\

\begin{table}[h]
\begin{center}
\centering \caption{Marginal changes of dropout rate in the mandatory vs. voucher system}
\renewcommand{\arraystretch}{1} \scriptsize \label{dropout}
\resizebox{160mm}{!}{\begin{tabular}{lcccccc}
\noalign{\smallskip}\hline\hline\noalign{\smallskip}	
\multicolumn{7}{l}{Dep. variable: Dropout yes/no}\\
& \multicolumn{2}{c}{(1)} & \multicolumn{2}{c}{(2)} & \multicolumn{2}{c}{(3)} \\ \hline\noalign{\smallskip}
Post-reform period &  -.047 & (.002) &  -.046 & (.002) & -.037 & (.002)  \\ \hline\noalign{\smallskip}
Personal characteristics & \multicolumn{2}{c}{No}   & \multicolumn{2}{c}{Yes}  & \multicolumn{2}{c}{Yes} \\
Education and occupation & \multicolumn{2}{c}{No}   & \multicolumn{2}{c}{Yes}  & \multicolumn{2}{c}{Yes} \\
Employment and welfare history & \multicolumn{2}{c}{No}   & \multicolumn{2}{c}{Yes}  & \multicolumn{2}{c}{Yes} \\
Timing of unemployment and programme start & \multicolumn{2}{c}{No} & \multicolumn{2}{c}{Yes} & \multicolumn{2}{c}{Yes} \\
State of residence & \multicolumn{2}{c}{No} & \multicolumn{2}{c}{Yes}  & \multicolumn{2}{c}{Yes} \\
Programme type and durations & \multicolumn{2}{c}{No}   & \multicolumn{2}{c}{No}   & \multicolumn{2}{c}{Yes} \\
\noalign{\smallskip}\hline\hline\noalign{\smallskip}
\end{tabular}}
\parbox{16cm}{Note: Marginal effects after probit estimations based on the sample of treated individuals in Sample A.}
\end{center}
\end{table}
	
We estimate different specifications in which we add more control variables. In column (3), we use all available controls variables including dummies for different planned course durations. In all specifications (1)-(3), the marginal effect of the time dummy on the dropout rate is significantly negative implying that the dropout rate decreases after the reform by about 4-5 percentage points. This supports our argumentation.\\
	
\clearpage

%%%%%%%%%%%%%%%%%%
%%% APPENDIX E %%%
%%%%%%%%%%%%%%%%%%

\section{Results for monthly earnings\label{app5}} \setcounter{table}{0} \setcounter{figure}{0}

\begin{figure}[h!]
\begin{center}
\caption{Overall reform, post-reform, and pre-reform treatment effects} \label{reform1}
\renewcommand{\baselinestretch}{1}
\begin{tabular}{c}
\includegraphics[width=10cm]{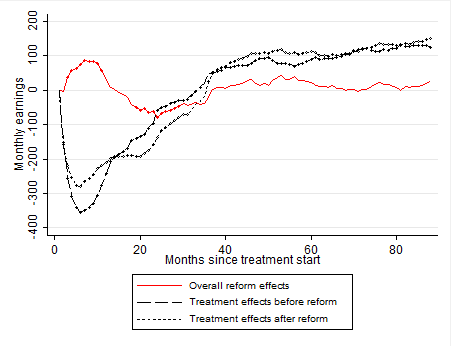} \\
\end{tabular}
\parbox{16cm}{\scriptsize Note: We estimate separate effects for each of the 88 months following the treatment. Diamonds indicate significant point estimates at the 5\%-level. Significance levels are bootstrapped with 499 replications. Lines without diamonds indicate point estimates that are not significantly different from zero. We use baseline Sample A and control for local employment agency district characteristics and the full set of observed characteristics (see Table \ref{balance} in Online Appendix \ref{app1}).}
\end{center}
\end{figure}

\begin{figure}[h!]
\begin{center}
\caption{Selection and overall reform effects}\label{reform2}
\renewcommand{\baselinestretch}{1}
\begin{tabular}{c}
\includegraphics[width=10cm]{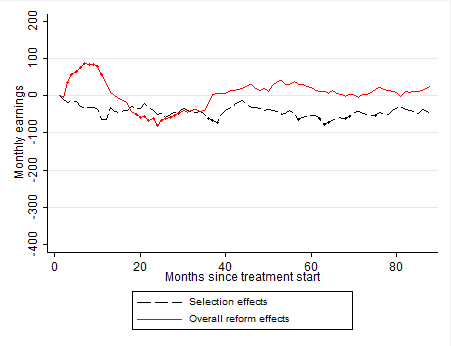} \\
\end{tabular}
\parbox{16cm}{\scriptsize Note: We estimate separate effects for each of the 88 months following the treatment. Diamonds indicate significant point estimates at the 5\%-level. Significance levels are bootstrapped with 499 replications. Lines without diamonds indicate point estimates that are not significantly different from zero. We use baseline Sample A and control for local employment agency district characteristics and the full set of observed characteristics (see Table \ref{balance} in Online Appendix \ref{app1}).}
\end{center}
\end{figure}

\setcounter{subfigure}{0}
\begin{figure}[h!]
\begin{center}
\caption{Time effects}\label{reform3}
\renewcommand{\baselinestretch}{1}
\begin{tabular}{c}
\includegraphics[width=10cm]{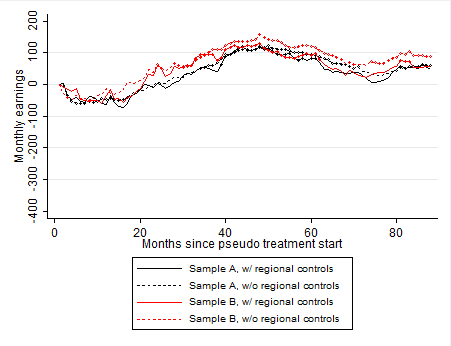} \\
\end{tabular}
\parbox{16cm}{\scriptsize Note: We estimate separate effects for each of the 88 months following the treatment. Diamonds indicate significant point estimates at the 5\%-level. Significance levels are bootstrapped with 499 replications. Lines without diamonds indicate point estimates that are not significantly different from zero. We use baseline Sample A and control for local employment agency district characteristics and the full set of observed characteristics (see Table \ref{balance} in Online Appendix \ref{app1}).}
\end{center}
\end{figure}

\begin{figure}[h!]
\begin{center}
\caption{Policy effects}\label{reform5}
\renewcommand{\baselinestretch}{1}
\begin{tabular}{c}
\includegraphics[width=10cm]{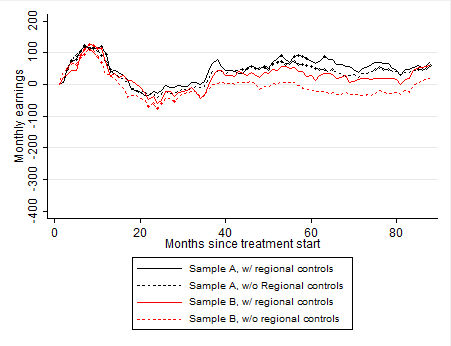} \\
\end{tabular}
\parbox{16cm}{\scriptsize Note: We estimate separate effects for each of the 88 months following the treatment. Diamonds indicate significant point estimates at the 5\%-level. Significance levels are bootstrapped with 499 replications. Lines without diamonds indicate point estimates that are not significantly different from zero. We use baseline Sample A and control for local employment agency district characteristics and the full set of observed characteristics (see Table \ref{balance} in Online Appendix \ref{app1}).}
\end{center}
\end{figure}

\setcounter{subfigure}{0}
\begin{figure}[h!]
\begin{center}
\caption{Decomposition of policy effect into course composition (indirect) and voluntary participation (direct) effects} \label{out2}
\renewcommand{\baselinestretch}{1}
\begin{tabular}{cc}
\subfigure[Sample A]{\includegraphics[width=7.5cm]{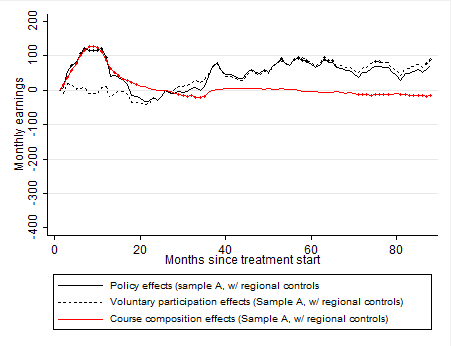}} & \subfigure[Sample B]{\includegraphics[width=7.5cm]{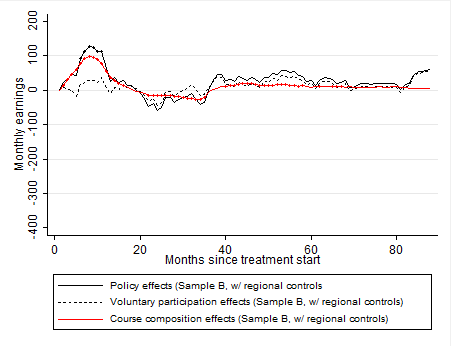}} \\
\end{tabular}
\parbox{16cm}{\scriptsize Note: We estimate separate effects for each of the 88 months following the treatment. Diamonds indicate significant point estimates at the 5\%-level. Significance levels are bootstrapped with 499 replications. Lines without diamonds indicate point estimates that are not significantly different from zero. We use baseline Sample A and control for local employment agency district characteristics and the full set of observed characteristics (see Table \ref{balance} in Online Appendix \ref{app1}). In the duration effects, we account for the planned course durations and interactions using fixed duration dummies.}
\end{center}
\end{figure}

\clearpage
\end{appendix}
\end{document}